\documentclass[a4paper,twocolumn,11pt,accepted=2025-03-16]{quantumarticle}
\pdfoutput=1
\usepackage[utf8]{inputenc}
\usepackage[english]{babel}
\usepackage[T1]{fontenc}
\usepackage{amsmath}
\usepackage{physics}
\usepackage{hyperref}
\usepackage{tikz}
\usepackage{lipsum}
\usepackage{float}
\usepackage{graphicx}
\usepackage{latexsym}
\usepackage{amssymb}
\usepackage{nicefrac}
\usepackage{amsfonts}
\usepackage{enumerate}
\usepackage{upgreek}
\usepackage{subfigure}
\usepackage{cite}
\usepackage{bm}
\usepackage[version=3]{mhchem}
\usepackage{nicefrac}
\usepackage{bbold}
\usepackage{verbatim}
\usepackage{multirow}
\usepackage{color}
\usepackage{comment}
\usepackage{xcolor}
\usepackage{soul}
\usepackage{placeins}
\usepackage{csquotes}

\begin{document}

\title{Transformer neural networks and quantum simulators: a hybrid approach for simulating strongly correlated systems}

\author{Hannah Lange}
\affiliation{Ludwig-Maximilians-University Munich, Theresienstr. 37, Munich D-80333, Germany}
\affiliation{Max-Planck-Institute for Quantum Optics, Hans-Kopfermann-Str.1, Garching D-85748, Germany}
\affiliation{Munich Center for Quantum Science and Technology, Schellingstr. 4, Munich D-80799, Germany}

\author{Guillaume Bornet}
\affiliation{Universit\'e Paris-Saclay, Institut d’Optique Graduate School,
CNRS, Laboratoire Charles Fabry, 91127 Palaiseau Cedex, France}

\author{Gabriel Emperauger}
\affiliation{Universit\'e Paris-Saclay, Institut d’Optique Graduate School,
CNRS, Laboratoire Charles Fabry, 91127 Palaiseau Cedex, France}

\author{Cheng Chen}
\affiliation{Universit\'e Paris-Saclay, Institut d’Optique Graduate School,
CNRS, Laboratoire Charles Fabry, 91127 Palaiseau Cedex, France}

\author{Thierry Lahaye}
\affiliation{Universit\'e Paris-Saclay, Institut d’Optique Graduate School,
CNRS, Laboratoire Charles Fabry, 91127 Palaiseau Cedex, France}

\author{Stefan Kienle}
\affiliation{FORRS Partners GmbH, Happelstr. 11, 69120 Heidelberg, Germany}

\author{Antoine Browaeys}
\affiliation{Universit\'e Paris-Saclay, Institut d’Optique Graduate School,
CNRS, Laboratoire Charles Fabry, 91127 Palaiseau Cedex, France}

\author{Annabelle Bohrdt}
\affiliation{Munich Center for Quantum Science and Technology, Schellingstr. 4, Munich D-80799, Germany}
\affiliation{University of Regensburg, Universitätsstr. 31, Regensburg D-93053, Germany}

\begin{abstract}
Owing to their great expressivity and versatility, neural networks have gained attention for simulating large two-dimensional quantum many-body systems. However, their expressivity comes with the cost of a challenging optimization due to the in general rugged and complicated loss landscape. Here, we present a hybrid optimization scheme for neural quantum states (NQS), involving a data-driven pretraining with numerical or experimental data and a second, Hamiltonian-driven optimization stage. By using both projective measurements from the computational basis as well as expectation values from other measurement configurations such as spin-spin correlations, our pretraining gives access to the sign structure of the state, yielding improved and faster convergence that is robust w.r.t. experimental imperfections and limited datasets. We apply the hybrid scheme to the ground state search for the 2D transverse field Ising model and dipolar XY model on $6\times 6$ and $10\times 10$ square lattices with a patched transformer wave function, using numerical data as well as experimental data from a programmable Rydberg quantum simulator [Chen et al., Nature 616 (2023)], and show that the information from a second measurement basis highly improves the performance. Our work paves the way for a reliable and efficient optimization of neural quantum states.
\end{abstract}

\maketitle

One of the key challenges in quantum many-body physics is the simulation of large, strongly correlated systems of more than one dimension. For these systems, exact parameterizations are unfeasible due to the exponential growth of the Hilbert space with system size. To overcome this problem, variational methods like tensor networks \cite{ORUS2014117} and their efficiently contractable variant, matrix product states (MPS)\cite{SCHOLLWOCK201196}, assume a certain form of the wave function, and this trial state is then optimized to obtain the best possible representation. However, in practice these methods face limitations, e.g. MPS are efficient parameterizations of states with area-law-entanglement \cite{Eisert2010,Hastings_2007}, but require an exponential number of parameters for states with volume-law entanglement. \\

Their great expressive power \cite{Cybenco1989,HORNIK1991251} and their flexibility make neural networks promising candidates to overcome these limitations. Recent works have shown that neural quantum states (NQS), i.e. variational wave functions parameterized by neural network parameters $\vec{\theta}$, are appealing trial states to model a broad range of quantum many-body systems, and even states with volume-law entanglement
\cite{Carleo2017,lange2024architectures,medvidović2024neuralnetwork}. In particular, transformer quantum states \cite{vaswani2017attention,zhang_transformer_2023,sprague2023variational,viteritti_transformer_2023,rende_optimal_2023,rende_simple_2023,Luo2022,Luo2023} have proven remarkable capabilities for ground state representations of spin systems up to system sizes of $10\times 10$, or, using a combination of recurrent neural networks and transformers, up to $16\times 16$.

\begin{figure}[t]
\centering
\includegraphics[width=0.48\textwidth]{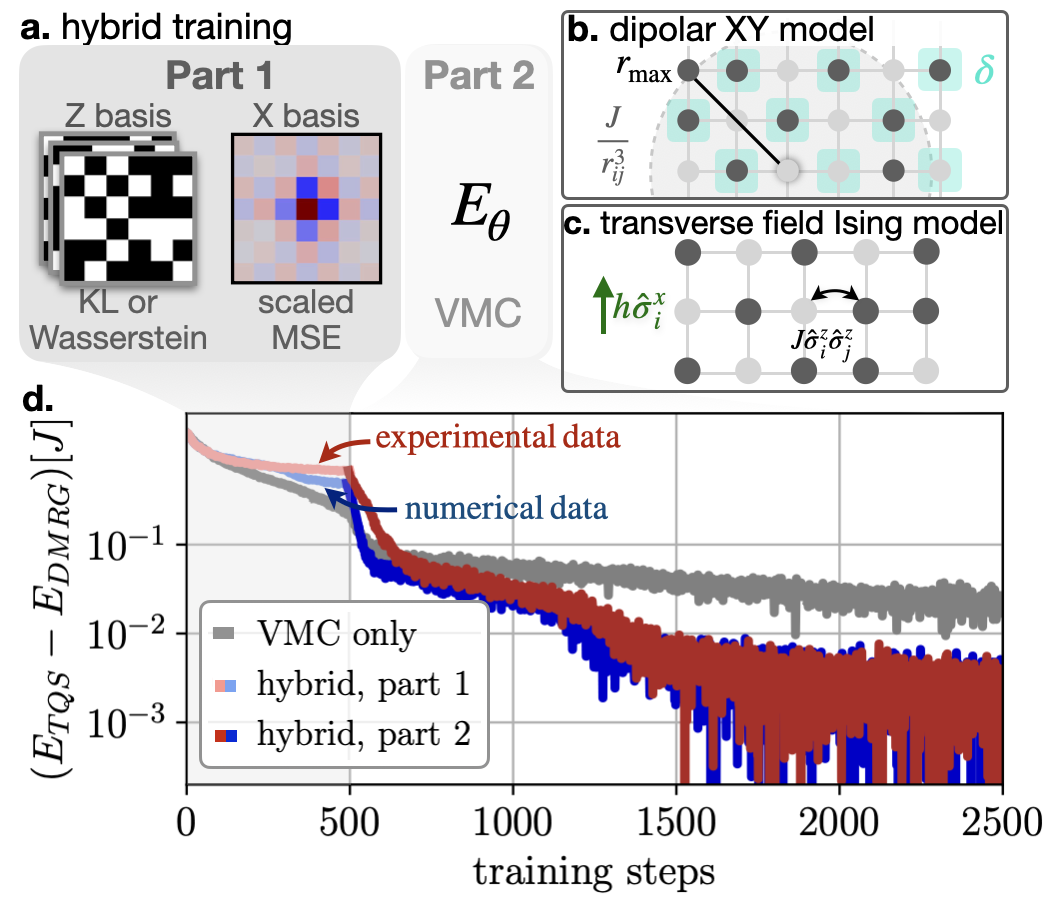}
\caption{ The hybrid training: \textbf{a.} We use a data-driven pretraining (part 1) based on the Wasserstein distance or the Kullback Leibler divergence (KL) for snapshots from the $Z$ basis and on the scaled mean-square error (MSE) for expectation values from the $X$ basis such as $\langle \hat{\sigma}^x_\mathbf{i}\hat{\sigma}^x_\mathbf{j}\rangle $ or $\langle \hat{\sigma}^x_\mathbf{i}\rangle $, before the Hamiltonian-driven variational Monte Carlo training (part 2). We apply this hybrid training scheme to the dipolar XY model (\textbf{b.}) with exchange interactions $J$ and a light shift $\delta$, and the transverse field Ising model (\textbf{c.}) with Ising interactions $J$ and a transverse field $h$. \textbf{d.} The hybrid scheme applied for a transformer quantum state with pretraining on Wasserstein and MSE loss improves the optimization result compared to the only VMC based training (gray lines), here for the dipolar XY model with $\hbar \delta/J \approx 0.004$.  } 
\label{fig:Fig1}    
\end{figure}

In most cases, the ground state search with NQS is started from randomly initialized parameters $\vec{\theta}_0$. If the optimization procedure converges, the trained NQS does not depend on $\vec{\theta}_0$. However, due to the in general very complicated loss landscape with many local minima, converging NQS often becomes very challenging \cite{Bukov2021,Park2020,Inack2022}. In these cases, the convergence and the convergence time can crucially depend on $\vec{\theta}_0$. Here, we use a data-driven pretraining from a programmable Rydberg quantum simulator or, to test our procedure with ideal data, from density-matrix renormalization group (DMRG) simulations, yielding a parameterization $\vec{\theta}_0$ that approximates the training data, similar to a quantum state reconstruction \cite{Torlai2018,Lohani_2020,Koutny2022}. The underlying idea is that any data that is relatively close to the target state, e.g. experimental data or (unconverged) numerical data, contains enough information about the actual system such that $\vec{\theta}_0$ is close to the ground state in the optimization landscape. After this first stage, the NQS is further optimized using a Hamiltonian-driven training procedure by employing variational Monte Carlo (VMC), see Fig.~\ref{fig:Fig1}a. Recent works have shown that this procedure can improve the convergence for stoquastic spin systems and molecular Hamiltonians \cite{czischek_data-enhanced_2022,moss_enhancing_2023,Bennewitz2022}, even in the presence of experimental imperfections and finite temperatures. 

In contrast to the stoquastic Hamiltonians considered in Refs. \cite{czischek_data-enhanced_2022,moss_enhancing_2023}, in general ground states have a sign structure and hence information from different measurement configurations than the computational ($Z$) basis is required. In particular, since optimizing the sign structure with variational Monte Carlo is in general challenging, information on the sign structure can drastically improve the performance, and is often included by hand as e.g. the Marshall sign rule for the Heisenberg model \cite{hibat-allah_recurrent_2020}. However, this is in general not possible since the sign structure is unknown. In these cases, information on the signs can be inferred from measurements outside the $Z$ basis. Since the computational cost of such a state reconstruction away from the $Z$ basis scales exponentially with the number of sites that are rotated outside the computational basis, in Refs. \cite{Bennewitz2022,Torlai2018} only snapshots with a few rotated sites are considered. \\

Here, we present two extensions to existing hybrid training schemes: $(i)$ We show that an efficient way for using information from other bases than the computational basis are the local magnetization or correlation maps, as shown in Fig.~\ref{fig:Fig1}a (dark gray box, right). This has the following advantages compared to training on snapshots away from the computational basis: a lower computational cost (e.g. for spin-spin correlations only two spins at a time have to be rotated away from the $Z$ basis); the possibility of reducing experimental imperfections by e.g. explicitly employing spatial symmetries; and the applicability of experimental or numerical data from methods that do not give access to snapshots such as Determinant Quantum Monte Carlo. $(ii)$ For the reconstruction training in the computational basis, we introduce another distance measure than the KL divergence, namely the Wasserstein distance \cite{Villani2009} (\ref{fig:Fig1}a, dark gray box, left), with the advantage that it allows to include the notion of a physically inspired \textit{distance between snapshots} which is not considered when using the KL divergence. \\

We evaluate the performance of our hybrid training scheme on two models: the 2D transverse field Ising model (TFIM)
\begin{align}
    \hat{\mathcal{H}}_\mathrm{TFIM}= 4J \sum_{\langle ij\rangle} \hat{S}_i^z\hat{S}_j^z +2h \sum_{i} \hat{S}_i^x,
    \label{eq:HamTFIM}
\end{align}
and the 2D dipolar XY model, with a staggered magnetic field on sublattice $B$, see Fig.~\ref{fig:Fig1}b,
\begin{align}
    \hat{\mathcal{H}}_\mathrm{XY}= J \sum_{i<j}\frac{a^3}{r_{ij}^3}\left[ \hat{S}_i^+\hat{S}_j^- + \hat{S}_i^-\hat{S}_j^+\right]+\hbar \delta \sum_{i\in B}\hat{n}_i^\uparrow,
    \label{eq:Ham}
\end{align}
with $\hat{n}_i^\uparrow= \hat{S}_i^z + \frac{1}{2}$, the spin $1/2$ operators $\hat{S}_i^z$ and $\hat{S}_i^\pm= \hat{S}_i^x\pm i\hat{S}_i^y$ and the antiferromagnetic (ferromagnetic) interaction strength $J>0\,(J<0)$, where we restrict the interaction radius to nearest neighbors (TFIM) and $\vert i-j\vert <3$ (dipolar XY). The latter system was recently realized on a programmable Rydberg simulator \cite{Chen2023}, where additionally small van der Waals interactions are present, see Appendix~\ref{appendix:Data}, which are taken into account as well.  

In order to capture the long-range interactions of the dipolar XY model, we use a quantum state representation in terms of a transformer quantum state (TQS) \cite{vaswani2017attention,zhang_transformer_2023,sprague2023variational,viteritti_transformer_2023,rende_optimal_2023,rende_simple_2023,fitzek2024,Luo2022,Luo2023} (see Appendix~\ref{appendix:transformer}) with two separate output layers for phase $\phi_{i,\vec{\theta}}$ and amplitudes $p_{i,\vec{\theta}}$ at each site $i$, i.e. the total wave TQS function is given by
\begin{align}
\ket{\psi}_{\vec{\theta}}=\sum_{\vec{\sigma}}\mathrm{exp}(i\phi_{\vec{\theta}}({\sigma}))\sqrt{p_{\vec{\theta}}({\sigma})}\ket {\sigma},
\end{align}
with $\phi_{\vec{\theta}}({\sigma})=\sum_i \phi_{i,{\theta}}({\sigma})$ and $p_{i,{\theta}}({\sigma})=\Pi_i p_{i,\vec{\theta}}({\sigma}\vert {\sigma}_{<i})$ \cite{zhang_transformer_2023}.
This architecture involves a so-called attention mechanism with trainable weights for all-to-all correlations between the patches. Specifically, we use the autoregressive and patched version of the TQS \cite{dosovitskiy2021image}. \\

In the remainder of this paper we introduce the hybrid optimization scheme and test it using a TQS representation  for the 2D TFIM \eqref{eq:HamTFIM} and dipolar XY models \eqref{eq:Ham} on $6\times 6$ and $10\times 10$ lattices. In all cases, we find significantly improved results compared to an only Hamiltonian-driven training or a hybrid training with data from only a single basis. Furthermore, the TQS proves to be remarkably efficient in representing the ground states of these systems compared to MPS. \\

\section{Hybrid training} 
The hybrid training procedure that we use consists of two parts, see Fig.~\ref{fig:Fig1}: A data-driven training on external data, e.g. experimental measurements (part 1) and a Hamiltonian-driven training using variational Monte Carlo \cite{McMillan1965,huang2017accelerated} (part 2, see Appendix~\ref{appendix:VMC}). \\

\textit{Data-driven pretraining (part 1): }The goal of the data-driven pretraining is to obtain a parameterization of the NQS that is already close in parameter space to the ground state, and which then serves as the initialization for the second part, namely the Hamiltonian-driven training. 

For snapshots from the computational $Z$ basis, in previous works \cite{czischek_data-enhanced_2022,moss_enhancing_2023} the Kullback-Leibler (KL) divergence between the measurement distribution $q$ and the NQS amplitudes $p_{\vec{\theta}}=\vert \psi_{\vec{\theta}}\vert^2$ was used for the pretraining, i.e.
\begin{align}
    D_\mathrm{KL}(q\vert p_{{\vec{\theta}}}) = \sum_{i=1}^n
    q({\sigma}^q_i)\, \mathrm{log}\left(
    \frac{q({\sigma}^q_i)}{ p_{{\vec{\theta}}}({\sigma}^q_i)}\right)\,
    \label{eq:KL}
\end{align}
for an underlying dataset with measurements $\{\sigma^q_{i}\}_{i=1,\dots,n}$. Note that $D_\mathrm{KL}$ is asymmetric, i.e. $D_\mathrm{KL}(q\vert p_{{\vec{\theta}}})\neq D_\mathrm{KL}( p_{{\vec{\theta}}}\vert q)$.

This loss function however has no information on the physical system under consideration. In order to add this information to the pretraining, we aim to include the notion of a (1D) distance $\lVert \sigma-\sigma^\prime \rVert_E$ between snapshots $\sigma$ and $\sigma^\prime$. Here, we choose the diagonal part of the energy $\lVert \sigma-\sigma^\prime \rVert_E := E_\mathrm{diag}(\sigma, \sigma^\prime)=\vert \langle \sigma \vert \hat{\mathcal{H}}\vert \sigma \rangle -\langle \sigma^\prime \vert \hat{\mathcal{H}}\vert \sigma^\prime \rangle\vert $, indicating e.g. if two snapshots $\sigma$ and $\sigma^\prime$ are close to each other in the sense that they are connected by symmetries $\lVert \sigma-\sigma^\prime \rVert_E=0$, only a few spin flips $\lVert \sigma-\sigma^\prime \rVert_E\approx 0$ or many spin flips $\lVert \sigma-\sigma^\prime \rVert_E>0$, see also Appendix~\ref{appendix:Datatraining}. 
To incorporate this information into a probability distribution metric, we use the Wasserstein distance instead of the KL divergence. The Wasserstein distance measures the minimal cost required to transform one distribution into another. The cost is measured by the the probability mass, i.e. the probability assigned to specific points in the distribution, that must be moved from $q$ (with samples $\{\sigma^q_{i}\}_{i=1,\dots,n}$) to obtain $p_{{\vec{\theta}}}$ (with samples $\{\sigma^p_{j}\}_{j=1,\dots,m}$), or vice versa. 
Formally, the probability mass transport is described by a matrix $\gamma\in \mathbb{R}_+^{n\times m}$, with $\sum_i \gamma_{ij}=p_{{\vec{\theta}}}(\sigma^p_j)$ and $\sum_j \gamma_{ij}=q(\sigma^q_i)$, i.e. $\gamma$ moves all probability mass from $q$ to $p_{{\vec{\theta}}}$ and vice versa. In this notation, the Wasserstein loss is \cite{Ye2022,flamary2021pot}
\begin{align}
    W(q,p_{{\vec{\theta}}}) = \mathrm{min}_{\gamma } \left[ 
    \sum_{i,j} \gamma_{ij}\lVert \sigma^q_i-\sigma^p_j\rVert_E
    \right].
    \label{eq:wasserstein}
\end{align}
Since there are many ways $\gamma$ to move the probability mass, a optimal transport problem has to be solved for the evaluation of Eq. \eqref{eq:wasserstein}, which can be done efficiently in 1D with a computational cost of $\mathcal{O}(n\mathrm{log}(n))$ \cite{flamary2021pot}. Note that in contrast to $D_{KL}$, $W$ is symmetric.
\\

Training on computational basis measurements gives access to the amplitudes of the state under consideration. In general, information from other measurement configurations than the computational basis is needed to get a good approximation of the state under consideration, in particular its sign structure \cite{Huang2020,quek2018adaptive,Smith2021,Lange_2023}. However, calculating the KL or Wasserstein distance for a measurement configuration $\vec{b}$ in an arbitrary basis scales exponentially with the number of spins $r$ that are rotated out of the $Z$ basis since the NQS wave function $\psi_{\vec{\theta}}$ has to be rotated to $\psi_{\vec{\theta}}^\prime(\sigma^{\vec{b}})$ in order to use Eqs.~\eqref{eq:KL} and \eqref{eq:wasserstein}, see Appendix~\ref{appendix:Datatraining}. Here, instead of training on the measurement distribution, we calculate the local magnetization or spin-spin correlations in the rotated basis from the measured snapshots and compare it to the TQS expectation values. Specifically, we consider $\langle \hat{\sigma}^x_\mathbf{i}\rangle_{(\vec{\theta})}$ for the TFIM ($\langle \hat{\sigma}^x_\mathbf{i}\hat{\sigma}^x_\mathbf{j}\rangle_{(\vec{\theta})}$ for the dipolar XY model, with a fixed reference site $\mathbf{i}$ located in the middle of the system). This is done by rotating the spins at $r=1( 2)$ sites $\mathbf{i}$ (and $\mathbf{j}$ for the correlations), i.e. with the same (with a smaller or the same) computational cost as in Ref. \cite{Torlai2018}(\cite{Bennewitz2022}), but with a systematic way of choosing the sites $\mathbf{i}$ and $\mathbf{j}$. Furthermore, for many quantum simulation platforms or numerical methods, observables like spin correlations are much more natural to obtain.

Training on expectation values requires a different loss function: Instead of the KL divergence, we use a (scaled) mean-square error (MSE) loss, see Appendix~\ref{appendix:corr_training}, which does not incorporate information on the measurement statistics. Instead, it allows to compensate for systematic errors that are e.g. present in experimental data: Firstly, spatial symmetries can be explicitly applied even if not perfectly represented in each snapshot, by averaging over the expectation values obtained from applying the respective symmetry. Furthermore, some expectation values appear more short-ranged in the experimental data as in theory due to experimental imperfections arising e.g. from the state preparation or losses. This justifies to use an asymmetrically scaled MSE such that the NQS is forced to represent states with higher absolute expectation values than the experimental target rather than smaller ones. More information on the training data can be found in Appendix~\ref{appendix:Data}.\\

\textit{Hamiltonian-driven training (part 2): } In order minimize the energy of the system further in the second part of the hybrid training, we use variational Monte Carlo (VMC) \cite{becca_sorella_2017, McMillan1965,huang2017accelerated}, minimizing the expectation value of the energy
\begin{align}
\langle E_{\vec{\theta}}\rangle = \sum_{\vec{\sigma}}\vert \psi_{\vec{\theta}}(\sigma)\vert^2\, E^{\mathrm{loc}}_{\vec{\theta}} (\sigma),
\label{eq:E}
\end{align}
with the local energy $E^{\mathrm{loc}}_{\vec{\theta}} (\sigma)=\frac{\langle \sigma\vert\mathcal{H}\vert\psi_{\vec{\theta}}\rangle }{\langle \sigma \vert\psi_{\vec{\theta}}\rangle}$, see Appendix~\ref{appendix:VMC}.\\

\begin{figure}[t]
\includegraphics[width=0.49\textwidth]{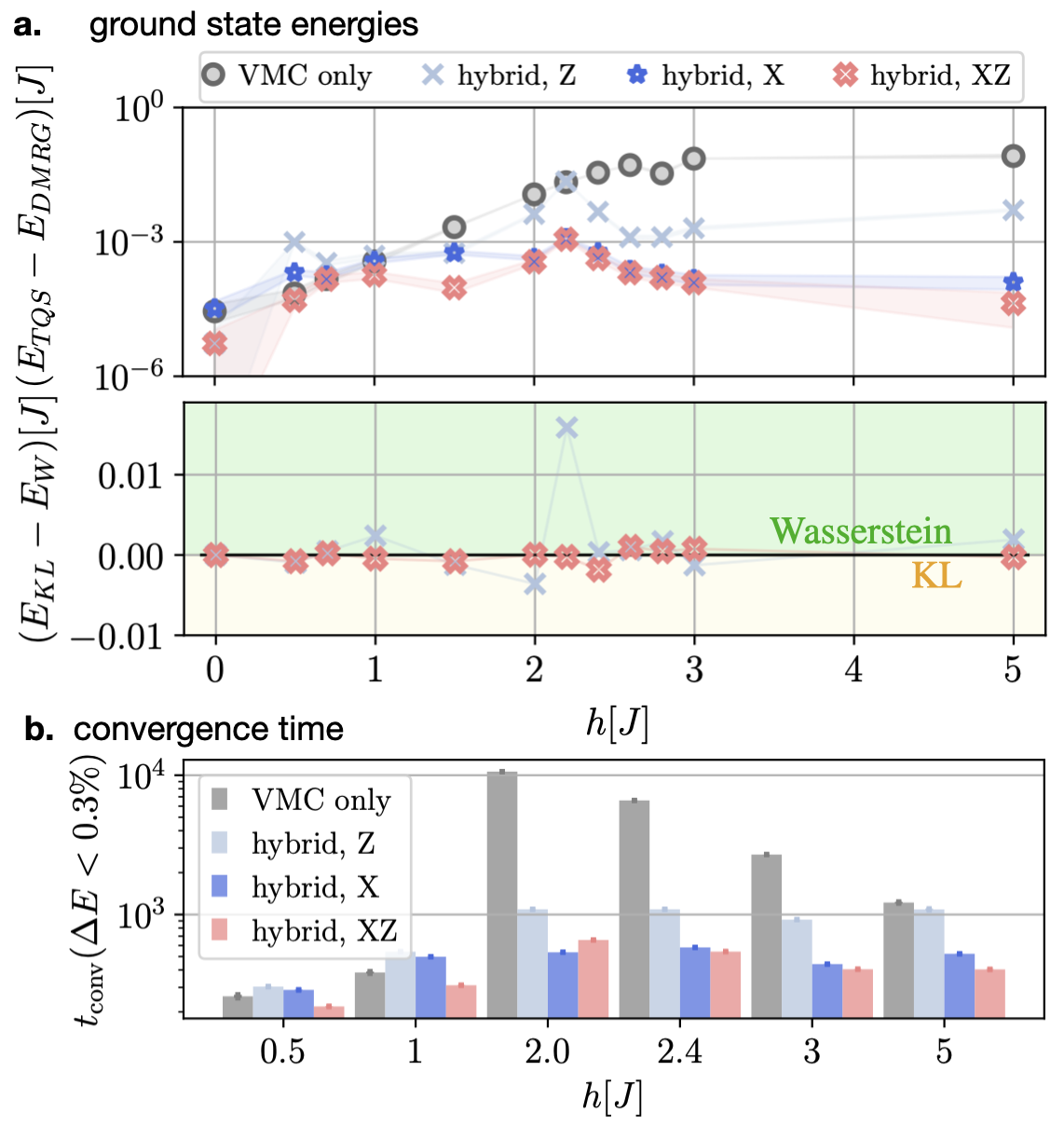}
\vspace{-0.3cm}
  \caption{Hybrid training of the 2D TFIM on a $6\times 6$ lattice using numerical data for the pretraining: \textbf{a.}~Top: Comparison of energy errors per site without pretraining (gray), with pretraining on data from only Z or X basis (blue) and on data from both bases (red) after $1000$ training steps. Bottom: Comparison of KL or Wasserstein loss for the snapshots from the $Z$ basis. We show the difference between $E_{KL}$ and $E_W$, for a training only in the $Z$ basis (blue) and in both bases (red). \textbf{b.}~Convergence time in units of training steps without and with pretraining on data from $Z$, $X$ and both bases, for which the energies averaged over $100$ epochs drop below $\Delta E = (E_{TQS}-E_{DMRG})/J<0.3\%$. }
  \label{fig:6x6_TFIM}
\end{figure}

\section{Results } 
The results for the 2D TFIM and the 2D dipolar XY model ground states on $6\times 6$ and $10\times 10$ lattices obtained with the hybrid training procedure are presented in Figs.~\ref{fig:6x6_TFIM} to \ref{fig:10x10}. Additional results and details on the architecture and the training can be found in Appendix~\ref{appendix:transformer}. We focus on antiferromagnetic (AFM) spin interactions in both cases. For ferromagnetic interactions, the convergence of the TQS starting from randomly initialized parameters is already very good, see Appendix~\ref{appendix:AddRes}, and hence not much room for improvement with the hybrid scheme. \\

\begin{figure}[t]
\includegraphics[width=0.49\textwidth]{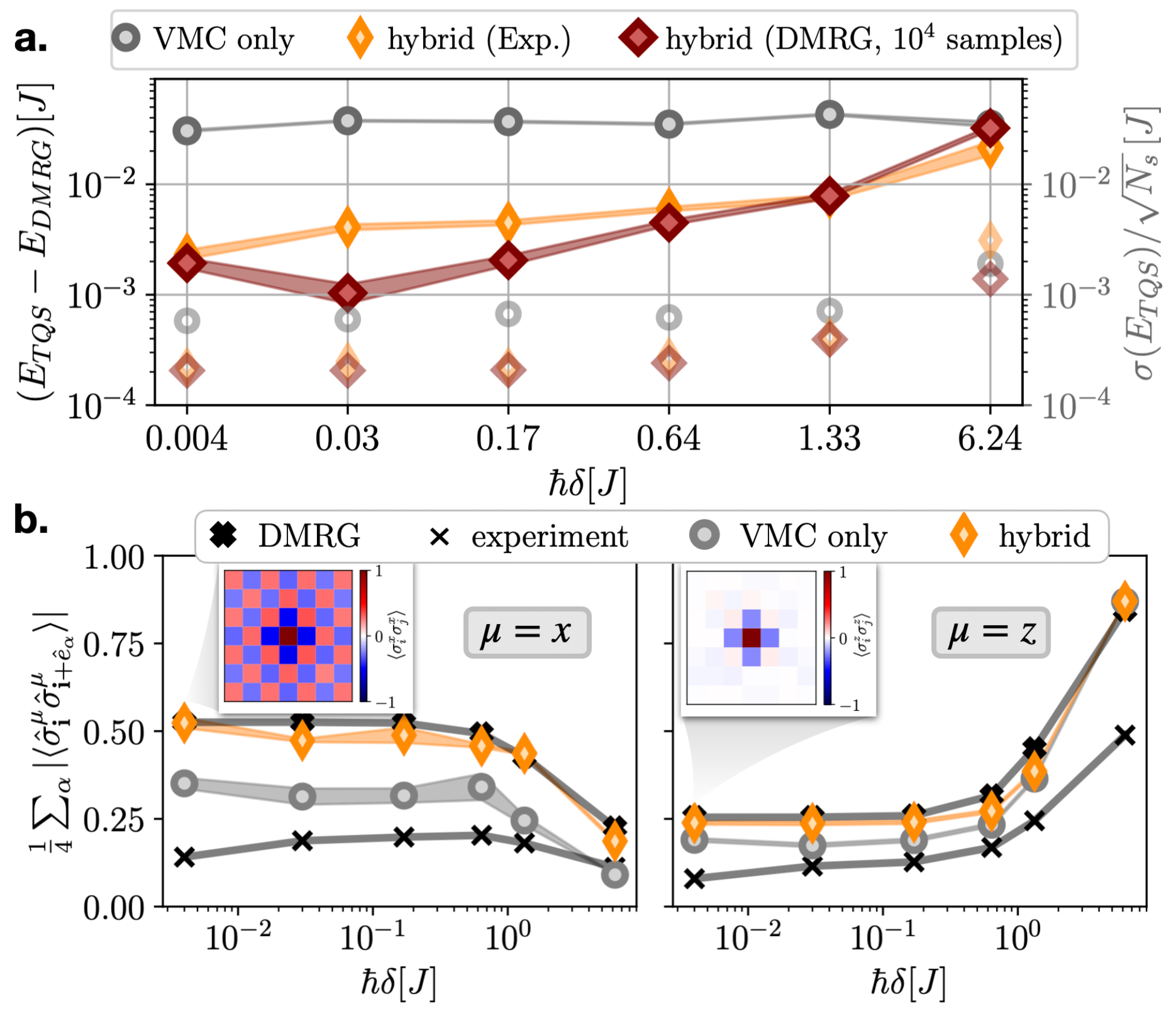}
  \caption{Hybrid training of the dipolar XY model on a $6\times 6$ lattice using experimental (Exp.) and numerical (DMRG) data for the pretraining: \textbf{a.}~Comparison of energy errors (solid lines) and standard deviation of the mean (light lines) obtained without pretraining (grey) and with pretraining on experimental (orange, $\approx 10^3$ samples) and numerical data (red, $10^4$ samples) after $2000$ training steps. \textbf{b.} The absolute correlations $\vert \langle \hat{\sigma}^\mu_\mathbf{i}\hat{\sigma}^\mu_{\mathbf{i}+\hat{e}_\alpha}\rangle \vert $ for $\mu=x,z$ (left, right) averaged over all nearest-neighbors $\alpha$ from DMRG and experiment (black) as well as the TQS with (without) pretraining using experimental data shown in orange (gray). Insets show $\langle \hat{\sigma}^\mu_\mathbf{i}\hat{\sigma}^\mu_{\mathbf{j}}$ for $\mathbf{i}=(3,3)$ in the middle of the system for $\hbar\delta/J=0.004$. }
  \label{fig:6x6}
\end{figure}

\textit{TFIM.-- } For the 2D TFIM, we use the KL divergence or Wasserstein loss for snapshots from the $Z$ basis and the scaled MSE for the local $X$ magnetization $\langle \hat{\sigma}_{\mathbf{i}}^x\rangle$ with $\langle \dots \rangle$ denoting the average over all snapshots. Both snapshots and local magnetization are obtained from ground state DMRG calculations. The results for a $6\times 6$ square lattice are shown in Fig.~\ref{fig:6x6_TFIM}. In the limit of a vanishing transverse field $h=0$, the system reduces to the Ising model, which we find to be very well approximated by the TQS with and without pretraining (Fig.~\ref{fig:6x6_TFIM}\textbf{a}). When turning on $h>0$, the accuracy of the TQS decreases without pretraining (gray lines). When pretraining the TQS on a single basis only (blue lines), the energy error can be reduced compared to the only VMC based training in some cases: In the regimes $J>h$ ($h>J$) the physics is dominated by the $Z$ ($X$) directions, which is reflected in a decreased error when training on the KL loss (scaled MSE) of data from these measurement configurations. In contrast, training on $Z$ ($X$) in the opposite regimes $J<h$ ($h<J$) does not improve the result as much. Our combined pretraining with data from $Z$ and $X$ basis in contrast significantly improves the results for all $h$. This is expected to be particularly relevant for more general systems where it is not known which basis is more relevant. 

Furthermore, we compare the results when using the KL divergence or the Wasserstein loss in Fig.~\ref{fig:6x6_TFIM}\textbf{a} (bottom), and find a similar improvement for both loss functions. 

Lastly, the hybrid scheme can tremendously speed up the calculations, see Fig.~\ref{fig:6x6_TFIM}\textbf{b}. For example, the convergence time (in units of the training steps) to obtain energy errors per site $(E_{TQS}-E_{DMRG})<0.003 J$ decreases by more than an order of magnitude when using the hybrid optimization for $h=2J$. Note that we have used the same hyperparameters for all cases, defined in Appendix, Tab.~\ref{tab:paramsTFIM}. \\


\begin{figure}[t]
\centering
\includegraphics[width=0.45\textwidth]{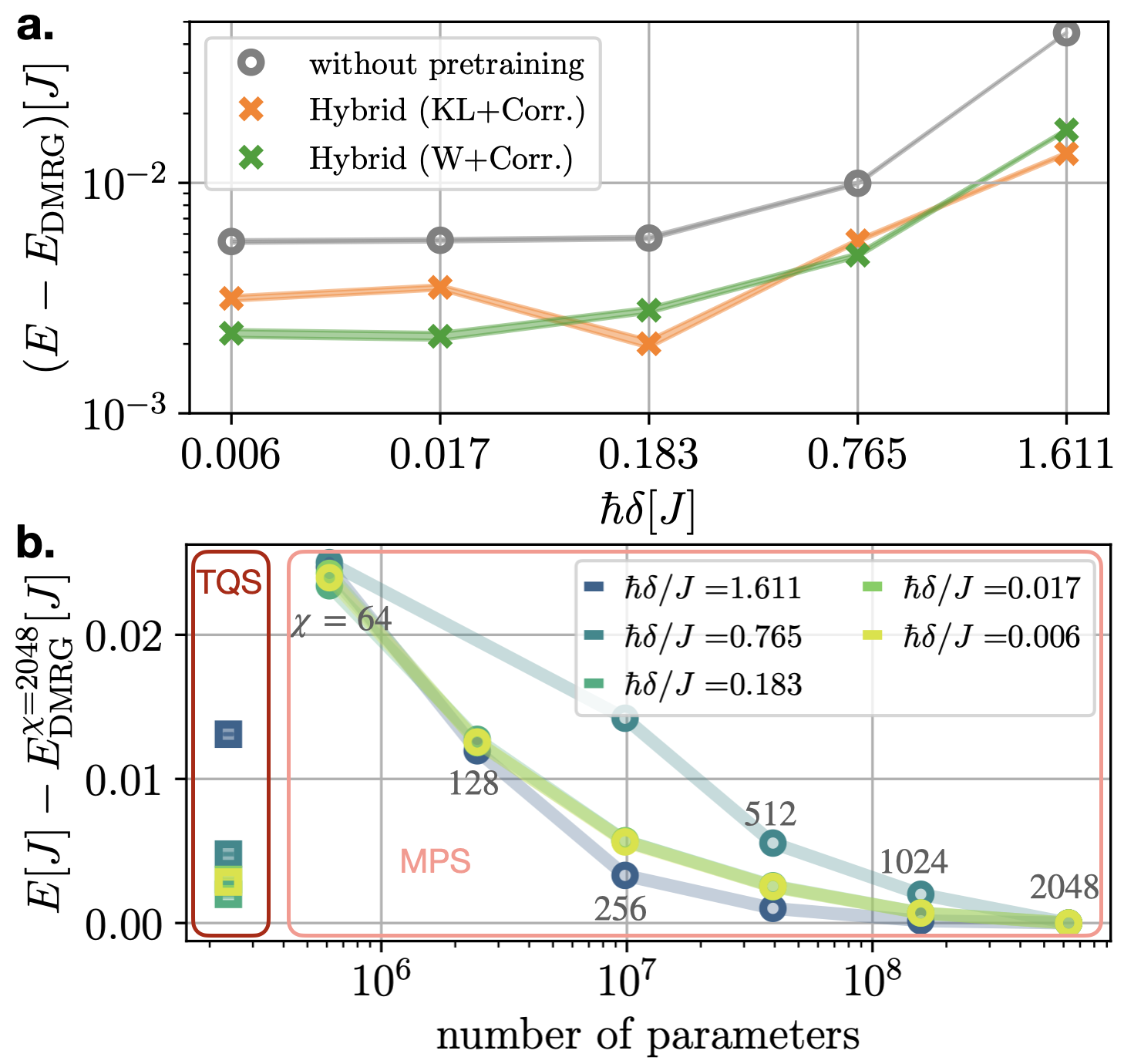}
\caption{Hybrid training of the dipolar XY model on a $10\times 10$ lattice using experimental data for the pretraining: \textbf{a.} Energy error per site (after $15,000$ training steps) obtained from only VMC training (gray lines) and hybrid training with pretraining on $\langle \hat{\sigma}^x_\mathbf{i}\hat{\sigma}^x_\mathbf{j}\rangle $ and the Kullback-Leibler divergence (Wasserstein distance) in orange (green) using around $10^2$ experimental measurements. \textbf{b.} Energy as function of the number of parameters used in the TQS (squares) and MPS (circles). For the MPS, we compare different bond dimensions $\chi=64\dots 2048$. } 
\label{fig:10x10}    
\end{figure}

\textit{Dipolar XY model.-- } For the dipolar XY model, we use the KL divergence / Wasserstein loss for snapshots from the $Z$ basis and the scaled MSE for spin-spin correlations $\langle \hat{\sigma}_{\mathbf{i}}^x\hat{\sigma}^x_{\mathbf{j}}\rangle$ in the $X$ basis, with a fixed reference site $\mathbf{i}$ located in the middle of the system. The experimentally obtained spin-spin correlations are averaged over different translations of the system, explicitly employing the translational invariance of the Hamiltonian under consideration. The results are shown in Figs.~\ref{fig:6x6} and \ref{fig:10x10} for $6\times 6 $ and $10\times 10 $ square lattices respectively. For both systems, we find significantly improved results when applying the hybrid learning scheme. This is apparent in terms of the energy per site (Figs.~\ref{fig:6x6}\textbf{a} and \ref{fig:10x10}\textbf{a}), its standard deviation of the mean (Fig.~\ref{fig:6x6}\textbf{a}) and observables such as the averaged nearest-neighbor spin-spin correlations in $X$ and $Z$ bases (Fig.~\ref{fig:6x6}\textbf{b}). Again, we use the same hyperparameters for our TQS in all cases, see Appendix, Tab.~\ref{tab:params}. Note that the energy error becomes large for $\hbar\delta>J$ for both system sizes, since the VMC optimization struggles with the fact that configurations which do not follow the checkerboard pattern of sublattices $A$ and $B$ defined by Eq. \eqref{eq:Ham} get a very large energy penalty, resulting in high variances during the optimization. This problem occurs for the optimization with and without pretraining. 

Furthermore, Fig.~\ref{fig:6x6}\textbf{a} shows that the quality of the data does not significantly influence the outcome of the hybrid learning scheme: the energy error per site is improved by the same order of magnitude when using numerical or experimental data, although experimental imperfections are clearly visible e.g. in terms of the spin-spin correlations in Fig.~\ref{fig:6x6}\textbf{b} (black lines). Even for the $10\times 10$ system, where only $87-204$ experimental snapshots for each $\delta$ were available, the pretraining improves the performance (Fig.~\ref{fig:10x10}\textbf{a}), showing the potential of hybrid training schemes even for very limited data sets.\\

With the improvement through hybrid training, the challenge of training neural quantum states is reduced and allows to make use of the powerful advantage of these ansätze: their great expressivity. This becomes apparent in Fig.~\ref{fig:10x10}\textbf{b}, where we show the energy error per site for the TQS and MPS ansätze w.r.t. the number of parameters that are used to encode the quantum state. It can be seen that the TQS is extremely efficient and uses around two orders of magnitude fewer parameters for the same energy error compared to the MPS representation at intermediate $\delta$.\\

\section{Summary and Outlook}
In this work, we have presented a hybrid approach which combines two powerful approaches for simulating quantum many-body systems: quantum simulators and neural quantum states. The hybrid scheme that we propose uses projective measurements and observables like spin-spin correlations from a quantum simulation platform to pretrain neural quantum states, before switching to the Hamiltonian-driven optimization in terms of variational Monte Carlo. 

In the first data-driven stage, we propose two innovations: $(i)$ For non-stoquastic Hamiltonians, information beyond the computational basis is necessary to extract the sign structure of the state, and we have demonstrated its importance for high-quality pre-training, see Fig.~\ref{fig:6x6_TFIM}. Since global rotations of the neural quantum state are exponentially costly, we train on site-resolved observables such as spin-spin correlations in the rotated basis. This comes with the advantage that certain experimental imperfections can be compensated, e.g. by explicitly applying spatial symmetries to the experimentally obtained expectation values before the pretraining. $(ii)$ For measurements in the computational basis, we propose to use the Wasserstein distance as the loss function instead of the Kullback-Leibler divergence, with the advantage that information on the Hamiltonian under consideration, in the form of a \textit{distance between snapshots}, is introduced to the data-driven training. \\

Our results on the 2D TFIM and the 2D dipolar XY model on $6\times 6$ and $10\times 10$ square lattices, testing pretrainings with numerical and experimental data, show that a combination of both $Z$ and $X$ measurement configurations is crucial to improve the optimization results and speed up the convergence over the full range of Hamiltonian parameters. Remarkably, this can be achieved even for very small data sets. Our work paves the way for an efficient and reliable simulation of more complicated systems such as fermionic quantum states or quantum spin liquids with NQS and quantum co-processors, using data from more than two measurement configurations as well as the Wasserstein distance for providing information on the Hamiltonian under consideration to the data-driven pretraining. For these systems, we envision a data processing step before the training to determine the relevant observables for the pretraining away from the computational basis \cite{Bohrdt2021,Verdel2024}.
\\


\emph{Code availability.--} The code used for this work is provided here: \url{https://github.com/HannahLange/HybridTransformer}. The implementation of the Wasserstein loss is build on the python package \cite{flamary2021pot}.\\

\emph{Note added.--} During the finalization of this manuscript, we became aware of Ref. \cite{fitzek2024}. In this work, an autoregressive, real-valued transformer wave function is trained on numerical data from the computational basis to represent the Rydberg Hamiltonian ground state for various system sizes and different points in the phase diagram.\\

\emph{Acknowledgements.--}
We thank Abhinav Suresh, Agnes Valenti, Christopher Roth, David Fitzek, Ejaaz Merali, Estelle Inack, Fabian Döschl, Fabian Grusdt, Henning Schlömer, Luciano Vitteriti, 
Lukas Homeier, Riccardo Rende, Roeland Wiersema, Roger Melko, Schuyler Moss, Tim Harris and Tizian Blatz for fruitful discussions. The data for the TFI models (see Appendix~\ref{appendix:TFIMExp}) were taken by P. Scholl, H. Williams and D. Barredo.
We acknowledge funding by the Deutsche Forschungsgemeinschaft (DFG, German Research Foundation) under Germany's Excellence Strategy -- EXC-2111 -- 390814868 and from the European Research Council (ERC) under the European Union’s Horizon 2020 research and innovation programm (Grant Agreement no 948141) — ERC Starting Grant SimUcQuam. The work at Institut d'Optique is supported by the Agence Nationale de la Recherche (ANR-22-PETQ-0004 France 2030, project QuBitAF), the European Research Council (Advanced grant No. 101018511-ATARAXIA), and the Horizon Europe programme HORIZON-CL4- 2022-QUANTUM-02-SGA (project 101113690 (PASQuanS2.1). HL acknowledges support by the International Max Planck Research School for Quantum Science and Technology (IMPRS-QST).

\section*{References}
\bibliographystyle{quantum}
\bibliography{main.bib}

\newpage~

\newpage~

\appendix

\renewcommand{\thefigure}{S\arabic{figure}}
\setcounter{figure}{0}
\onecolumngrid \section*{APPENDIX}

\section{Transformer wave function \label{appendix:transformer}}
\subsection{The architecture}
\begin{figure}[t]
\centering
\includegraphics[width=0.48\textwidth]{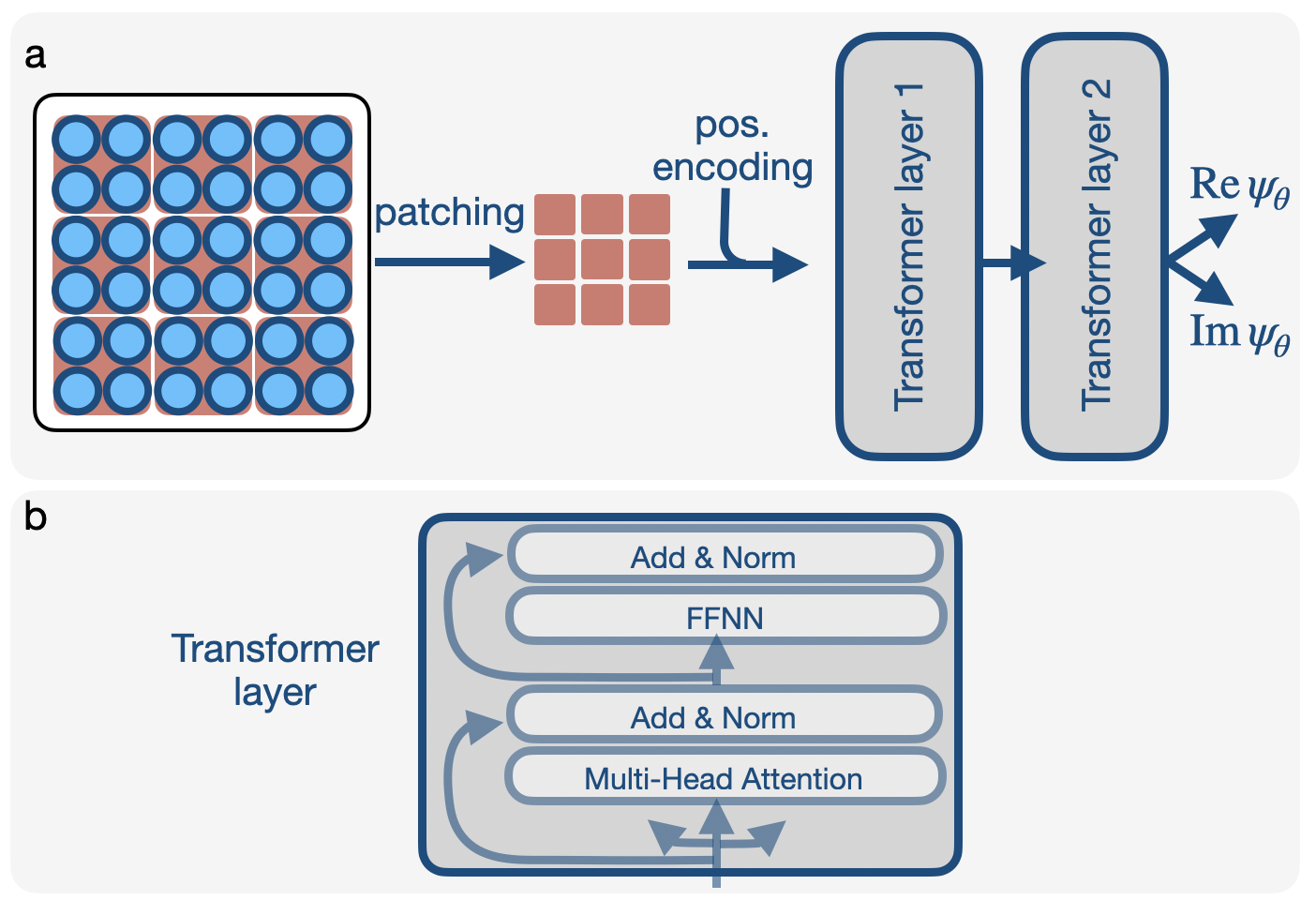}
\caption{The patched transformer architecture: \textbf{a.} Patches of $2\times 2$ sites are embedded and positional encoded, before being passed through two transformer layers, with separate output layers for amplitude and phase of the wave function. \textbf{b.} The transformer layer with multi-head attention, addition, normalization and feed forward neural network (FFNN). } 
\label{fig:Transformer}    
\end{figure}
Here, we use a representation of the dipolar XY model ground states $\ket{\psi}_{\vec{\theta}}$ in terms of a transformer neural network. Transformer networks are characterized by their attention mechanism \cite{vaswani2017attention}, see Fig.~\ref{fig:Transformer}b, consisting of trainable, all-to-all connections throughout the system, which allow the network to capture correlations between any sites in the system regardless of the position. 
When a local input configuration $\sigma_{i}$ at site $i$ is passed to a transformer layer, it is first embedded into a, typically higher dimensional, space of dimension $d_h$. Furthermore, a positional encoding is added, that encodes the position ${i}$. Then, the encoded input $\tilde{\sigma}_{i}$ is projected on a query vector ($q_i$), key vector ($k_i$) and a value vector ($v_i$), 
using trainable matrices $W^q, W^k, W^v \in \mathbb{R}^{d_h \times d_h}$. By defining query, key and value matrices as $Q = (q_1, ..., q_{N})$,  $K = (k_1, ..., k_{N})$ and  $V = (v_1, ..., v_{N})$, the self-attention mechanism is given by
\begin{align}
    \text{Attention}(Q, K, V) = \mathrm{softmax}\left(\frac{QK}{\sqrt{d_h / h}} + M\right)\,V.
\end{align}
Here, $h=1$ and $M$ is a \textit{mask} that is added to the signal, masking out all sites that $j>i$ and making the model autoregressive \cite{vaswani2017attention}. In a multi-headed attention mechanism each query, key and value vector is mapped to $h>1$ vectors with trainable weight matrices. In most cases, several transformer layers are stacked on top of each other.

Using two different output layers for phase and amplitude of the quantum state at each site, $\phi_{i,\vec{\theta}}(\vec{\sigma})$ and $p_{i,\vec{\theta}}(\vec{\sigma}\vert \vec{\sigma}_{<i})$ respectively, we define the transformer wave function by
\begin{align}
\ket{\psi}_{\vec{\theta}}=\sum_{\vec{\sigma}}\mathrm{exp}(i\phi_{\vec{\theta}}(\vec{\sigma}))\sqrt{p_{\vec{\theta}}(\vec{\sigma})}\ket {\sigma},
\end{align}
with $\phi_{\vec{\theta}}(\vec{\sigma})=\sum_i \phi_{i,\vec{\theta}}(\vec{\sigma})$ and $p_{i,\vec{\theta}}(\vec{\sigma})=\Pi_i p_{i,\vec{\theta}}(\vec{\sigma}\vert \vec{\sigma}_{<i})$
see also Ref. \cite{zhang_transformer_2023}. Note that by using a softmax activation function for the amplitude layer, both local and total amplitudes are normalized, making the TQS autoregressive. Furthermore, we apply rotational symmetry w.r.t. rotations by an angle of $\pi$, as detailed in Ref. \cite{Reh2023}.

The self-attention mechanism makes the transformer very expressive, but on the other hand also slows down the training, since every all-to-all connection has to be learned. Patched transformers, inspired from Vision transformers (ViT) \cite{dosovitskiy2021image}, overcome this problem by splitting the system into small patches, see Fig.~\ref{fig:Transformer}a. In Refs. \cite{rende_optimal_2023,rende_simple_2023,viteritti_transformer_2023,sprague2023variational}, the authors have shown that patched transformers can be used successfully for frustrated spin systems and Rydberg systems. \\

\subsection{Hyperparameters}
In this work, we use a patch size of $2\times 2$ and two transformer layers. The network parameters and training hyperparameters are listed in Tabs.~\ref{tab:paramsTFIM} and \ref{tab:params}. Before the training, we initialize the network parameters according to a uniform distribution between $-0.1$ and $0.1$ and set all biases of the output layers to zero. For the TFIM systems we use a exponential decay starting at epoch $n_\mathrm{start}$ with decay rate $\gamma$. For the dipolar XY model, we observe that gradients can become large and find it convenient to use a cosine learning rate schedule
\begin{align}
    l(n) = l_\mathrm{max} \cdot \mathrm{max}\left[ \eta \frac{1}{2} (1 + \mathrm{cos}(\pi \frac{n}{n_\mathrm{max}}),0.01\right]
    \label{eq:lrschedule}
\end{align}
with a prefactor $\eta=\begin{cases}
    \,1 \quad \mathrm{if} \, n > n_\mathrm{w}\\
    \frac{n}{n_\mathrm{w}} \quad \mathrm{else}
\end{cases}$ that reduces the learning rate in the warmup phase $n < n_\mathrm{w}$.
All benchmark calculations without pretraining are done with the same architectures and hyperparameters.

\begin{table}[htp]
\centering
\begin{tabular}{|l|c|}\hline
 & $6\times 6$  \\\hline
number of transformer layers & $2$  \\
hidden dimension $d_h$ & $32$  \\
FFNN dimension (see Fig.~\ref{fig:Transformer}) & $4d_h$  \\
total number of parameters & $28,800$  \\
pretraining learning rate & $5\cdot 10^{-4}$ \footnote{Except for the training on only $Z$ basis measurements and $h/J=5$, where the pretraining learning rate is decreased by a factor $1/5$ to avoid getting stuck in a local minimum.}   \\
VMC learning rate $l_\mathrm{max}$ & $ 2\cdot  10^{-3}$ \\
learning rate schedule $l(n)$ &  $n_\mathrm{start}=300$, $\gamma=0.9998$ \\
maximal no. of pretraining steps $n_\mathrm{pretrain}^\mathrm{max}$ & $100$\\
number of samples in VMC training (part 2) & $512$\\
number of samples used for the data points in Fig.~\ref{fig:6x6_TFIM} & $100\times 512$\\
errorbars in Fig.~\ref{fig:6x6_TFIM} & standard deviation of the mean\\
number of training steps used in Fig.~\ref{fig:6x6_TFIM} & $1,000$\\\hline
\end{tabular}
\caption{Transformer parameters and training hyperparameters for the $6\times 6$ 2D TFIM systems (Fig.~\ref{fig:6x6_TFIM}). We refer to the total number of training steps including the pretraining as $n$. Futhermore, we use a exponential decay rate starting at epoch $n_\mathrm{start}$ with decay rate $\gamma$.}
\label{tab:paramsTFIM}
\end{table}

\begin{table}[htp]
\centering
\begin{tabular}{|l|c|c|}\hline
 & $6\times 6$ & $10\times 10$ \\\hline
number of transformer layers & $2$  & $2$ \\
hidden dimension $d_h$ & $56$ & $88$ \\
FFNN dimension (see Fig.~\ref{fig:Transformer}) & $4d_h$ & $5d_h$ \\
total number of parameters & $85,320$ & $238,424$ \\
pretraining learning rate & $10^{-4}$   & $10^{-5}$  \\
VMC learning rate $l_\mathrm{max}$ & $ 5\cdot 10^{-4}$ & $10^{-4}$ \\
learning rate schedule $l(n)$ &  $n_\mathrm{max}=5,000$, $n_\mathrm{w}=1,000$&  $n_\mathrm{max}=20,000$, $n_\mathrm{w}=10,000$ \\
max. no. of pretraining steps $n_\mathrm{pretrain}^\mathrm{max}$ & $200$&$1,500$\\
number of samples in VMC training & $256$&$256$\\
number of samples in Figs.~\ref{fig:6x6} \& \ref{fig:10x10} & $10,000$ & $100\times 256$\\
errorbars in Figs.~\ref{fig:6x6} \& \ref{fig:10x10}  & standard dev. of the mean & standard dev. of the mean\\
no. of training steps in Figs.~\ref{fig:6x6} \& \ref{fig:10x10} & $2,000$ & $15,000$\\\hline
\end{tabular}
\caption{Transformer parameters and training hyperparameters for the $6\times 6$ and $10\times 10$ dipolar XY systems (Figs.~\ref{fig:6x6} and \ref{fig:10x10}). We refer to the total number of training steps including the pretraining as $n$ and use a cosine learnig rate schedule \eqref{eq:lrschedule}.}
\label{tab:params}
\end{table}

\FloatBarrier
\section{Data driven (pre-)training \label{appendix:Datatraining}}
\subsection{Measurements in computational basis: Kullback-Leibler divergence and Wasserstein distance} 

\begin{figure}[t]
\centering
\includegraphics[width=1\textwidth]{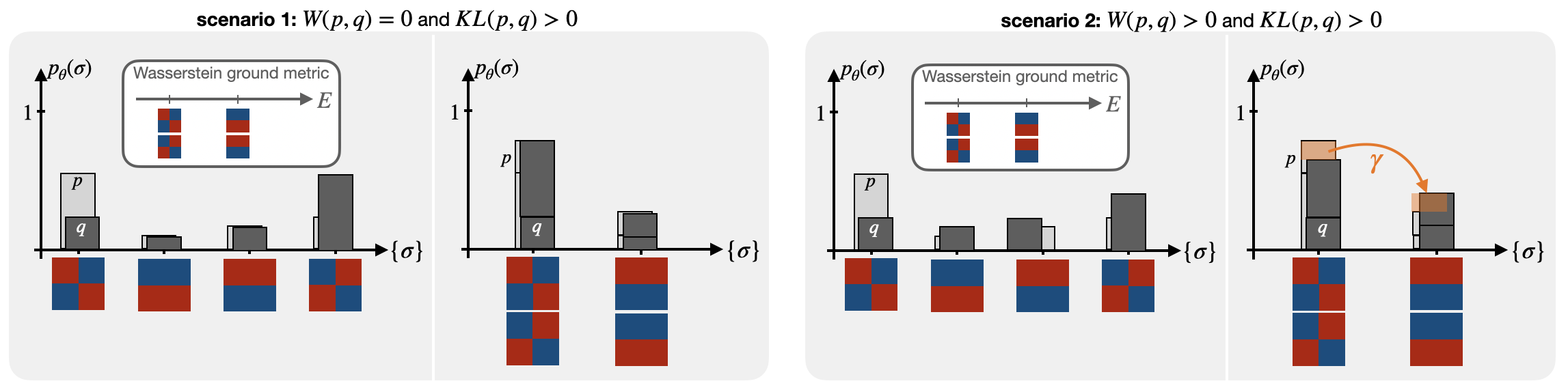}
\caption{Comparison between Kullback-Leibler (KL) divergence and Wasserstein distance (W) with energy as the ground metric for the exemplary case of a $2\times 2$ Heisenberg model. We focus on two scenarios: \textbf{scenario 1:} We show two distributions $q$ and $p$, each with peaks at different N\'eel configurations but otherwise same probability mass (left). Ordering the configurations according to their energy as done for the Wasserstein distance (right) yields a vanishing $W(p,q)=0$. In contrast $KL(p,q)\neq 0$ since it has no notion of the \enquote{closeness} (w.r.t. ground state energies) of the N\'eel states. \textbf{scenario 2:} For two completely different distributions $q$ and $p$, $W(p,q)$ measures the amount of probability mass that has to be transported from $q$ to $p$ or vice versa. Hereby, $\gamma$ is the transport matrix. For larger systems, there are many possible transport ways, and hence the optimal $\gamma$ has to be found, see Eq. \eqref{eq:wasserstein}.   } 
\label{fig:DistMeasures}    
\end{figure}
For the training on data from the computational basis, we compare two loss functions: The Kullback-Leibler (KL) divergence defined in Eq. \eqref{eq:KL} and the Wasserstein (W) distance defined in Eq. \eqref{eq:wasserstein}. The main advantage of the Wasserstein distance is that it incorporates the definition of a ground metric, capturing the distance between the snapshots in the training data set. Here, we use the diagonal part of the energy $E_\mathrm{diag}(\sigma^p, \sigma^q)=\vert \langle \sigma^p \vert \hat{\mathcal{H}}\vert \sigma^p \rangle -\langle \sigma^q \vert \hat{\mathcal{H}}\vert \sigma^q \rangle\vert $ as the ground metric. This implicitly captures the fact that configurations that are connected e.g. by a few spin flips, exchanges or certain rotations, are close to each other. The difference between the Kullback-Leibler divergence (KL) and the Wasserstein distance (W) are shown for two exemplary scenarios in Fig.~\ref{fig:DistMeasures}: 

\begin{itemize}
    \item \textbf{scenario 1:}  Fig.~\ref{fig:DistMeasures}a shows how the Wasserstein distance can incorporate the information on the Hamiltonian in the data-driven training. We show two distributions $q$ and $p$, each with peaks at different N\'eel configurations but otherwise same probability mass. For these distributions, we can calculate $KL(q,p)>0$. However, the KL does not take into account that for the Heisenberg model, the two N\'eel configurations and the two remaining configurations are energetically degenerate and can hance be treated as equivalent. In contrast, when calculating the Wasserstein distance, the configurations are ordered according to their (diagonal) energy $E$, resulting in the distributions shown in Fig.~\ref{fig:DistMeasures}b. In this exemplary scenario, it now becomes apparent that the distributions $p$ and $q$ are the same when degenerate configurations are grouped together, i.e. $W(q,p)=0$.
    \item \textbf{scenario 2:} Fig.~\ref{fig:DistMeasures}b shows two completely different distributions $q$ and $p$ (left). After ordering the configurations according to the (diagonal) energy (right), $W(p,q)$ measures the amount of probability mass that has to be transported from $q$ to $p$ or vice versa. Hereby, $\gamma$ is the transport matrix, as defined in Eq. \eqref{eq:wasserstein}. Since for larger systems, there are many possible transport ways $\gamma$, an optimal transport problem to find the optimal $\gamma$ has to be solved in each iteration. 
\end{itemize}

For both KL or the Wasserstein distance, we use the Adam optimizer \cite{kingma2017adam} for the minimization of the respective loss function.

\subsection{Training on data away from the computational basis}
\subsubsection{Exponential scaling w.r.t. spins rotated away from computational basis \label{appendix:rotatedprobs}}
Here, we review why evaluating the Kullback-Leibler divergence or the Wasserstein distance for snapshots away from the computational basis is exponentially costly. We consider a local rotation of $r$ spins from the $Z$ to the $X$ basis, using
\begin{align}
    \hat{U} = \hat{U}_1\otimes \hat{U}_2\otimes \dots \hat{U}_N,
\end{align}
with the local spin rotations given by 
\begin{align}
    \hat{U_i} = \frac{1}{\sqrt{2}}\begin{pmatrix}
1 & 1 \\
1 & -1 
\end{pmatrix} \quad \mathrm{or} \quad \hat{U_i} =1_{2\times 2}
\end{align}
at each site of the system. To evaluate the Kullback-Leibler divergence or the Wasserstein distance for snapshots in the global $X$ basis, $\sigma^x$, one needs  to calculate
\begin{align}
    \psi_{\vec{\theta}}^\prime(\sigma^x) = \langle \sigma^x\vert \hat{U}\vert \psi_{\vec{\theta}}\rangle = \sum_\sigma \langle \sigma^x \vert \hat{U}\vert \sigma\rangle \psi_{\vec{\theta}}(\sigma)= \sum_\sigma \langle \sigma^x \vert (\hat{U}_1\otimes \hat{U}_2\otimes \dots \hat{U}_N)\vert \sigma\rangle \psi_{\vec{\theta}}(\sigma).
\end{align}
For one rotated spin $r=1$, this involves the evaluation of $2^1$ terms, for $r=2$ it involves $2\cdot 2=2^2$ terms etc.. 

\subsubsection{Training on expectation values in the $X$ basis \label{appendix:corr_training}}
\begin{figure}[t]
\centering
\includegraphics[width=0.95\textwidth]{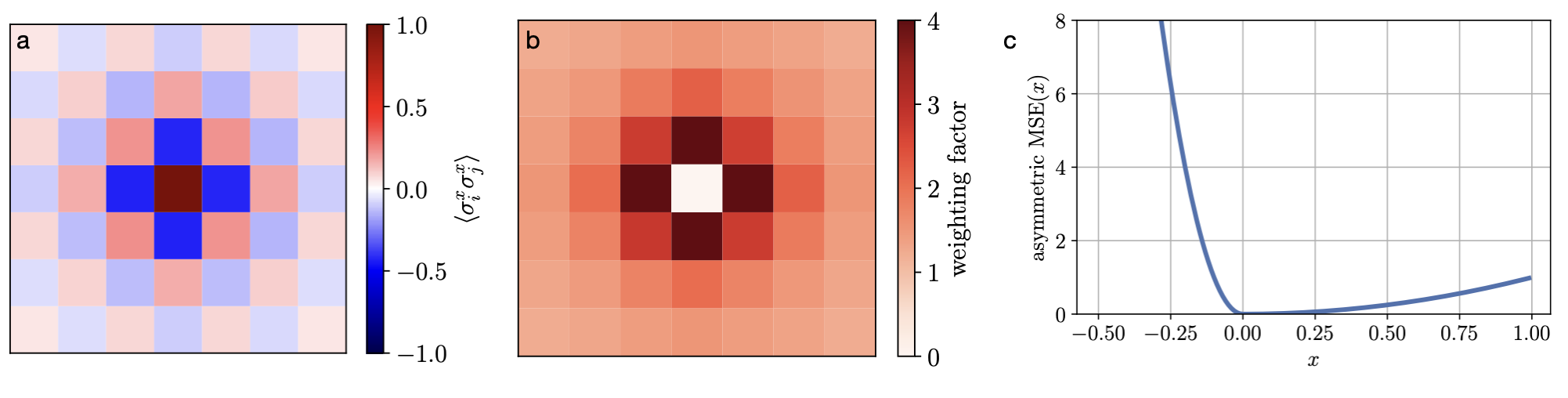}
\caption{Correlation-driven training in the $X$ basis by minimizing $\mathcal{C}=\sum_j \eta_j \cdot \mathrm{asymmetricMSE}(\langle \sigma^x_i \sigma^x_j\rangle_{\vec{\theta}} -\langle \sigma^x_i \sigma^x_j\rangle )$: \textbf{a.} Exemplary correlation map for $\delta\approx 1.35$. \textbf{b.} Weighting factor $\eta_j = 1/(1-\vert \langle \sigma^x_i \sigma^x_j\rangle \vert)^4 $. c) We use an asymmetric MSE with a higher penalty if $\vert \langle \sigma^x_i \sigma^x_j\rangle_{\vec{\theta}} \vert < \vert \langle \sigma^x_i \sigma^x_j\rangle \vert $, here shown for $x=(x_1,x_2)$ with $x_1,x_2\geq 0$.} 
\label{fig:scaledMSE}    
\end{figure}

The information from measurements away from the computational basis are incorporated by considering spin correlation maps instead of the measurement statistics. Specifically, we consider $\langle \hat{\sigma}^x_i\hat{\sigma}^x_j\rangle_{(\vec{\theta})}$ from the measurements (the NQS) and define the scaled mean-square error (MSE) loss,
\begin{align}
    \mathcal{C}=\sum_j \eta_j \cdot \mathrm{asymmMSE}(\langle \hat{\sigma}^x_i \hat{\sigma}^x_j\rangle_{\vec{\theta}},\langle \hat{\sigma}^x_i \hat{\sigma}^x_j\rangle ),
    \label{eq:scaledMSE}
\end{align}
where $i$ is fixed to the middle of the system and
\begin{align}
    \mathrm{asymmMSE}(x_1, x_2)=\begin{cases}
			f*(x_1-x_2)^2, & \text{if $\vert x_1\vert<\vert x_2\vert $ }\\
            (x_1-x_2)^2, & \text{else}
		 \end{cases},
\end{align}
$f>1$ is a factor that gives a higher penalty if $\vert \langle \sigma^x_i \sigma^x_j\rangle_{\vec{\theta}}\vert <\vert \langle \hat{\sigma}^x_i \hat{\sigma}^x_j\rangle\vert$ and $\eta_j = 1/(1-\vert \langle \sigma^x_i \sigma^x_j\rangle \vert)^4 $ weights the contributions to $\mathcal{C}$ depending on the magnitude of the target correlation value. Fig.~\ref{fig:scaledMSE} shows the weighting factor $\eta_j$ (b) for the example correlation map shown in a), as well as and the asymmetric MSE (c).

Hereby, evaluating 
\begin{align}
    \langle \hat{\sigma}^x_i\hat{\sigma}^x_j\rangle_{(\vec{\theta})} = \frac{1}{N_s}\sum_{\mathbf{x}}\sum_{\mathbf{x}^\prime}\frac{\psi_{\vec{\theta}}(\mathbf{x}^\prime)}{\psi_{\vec{\theta}}(\bf{x})}\langle \mathbf{x}\vert \hat{\sigma}^x_i\hat{\sigma}^x_j\vert \mathbf{x}^\prime\rangle
\end{align}
involves the computation of only $4$ terms. As for the computational basis, we use the Adam optimizer \cite{kingma2017adam} to optimize according to Eq. \eqref{eq:scaledMSE}.

\FloatBarrier
\section{Energy driven training \label{appendix:VMC}}
\begin{figure}[htp]
\centering
\includegraphics[width=0.8\textwidth]{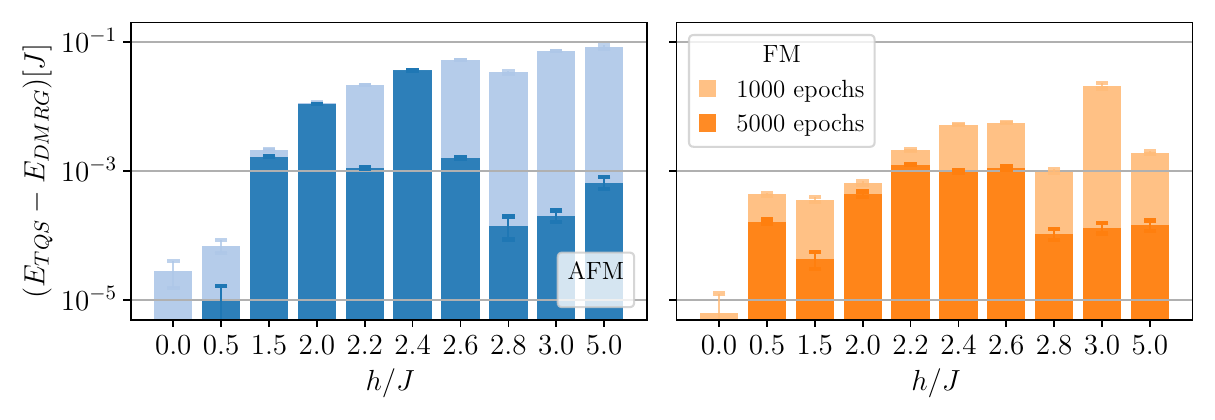}
\includegraphics[width=0.8\textwidth]{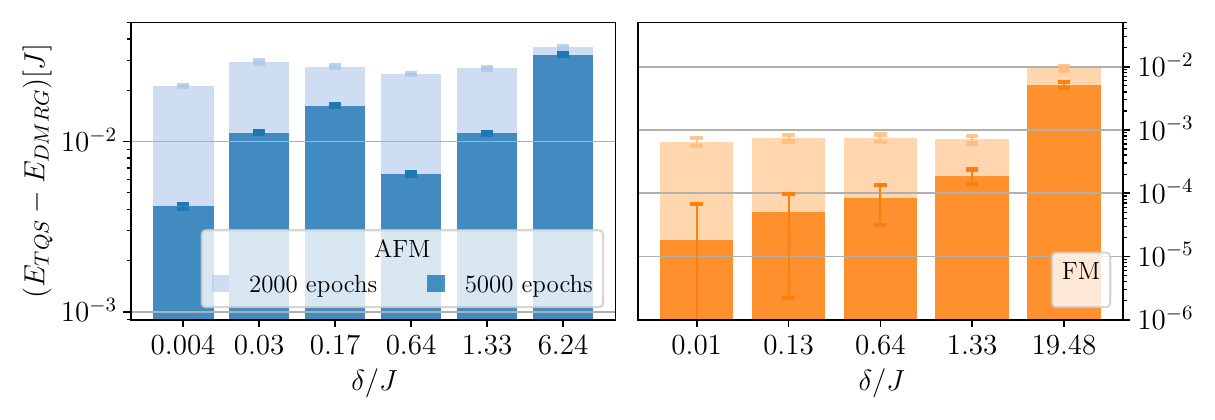}
\caption{Top: Transverse field ising model without pretraining. We show the energy error per site for $6\times 6$ AFM (left) and FM (right) TFIM after $1000$ and $5000$ epochs of only energy driven training. Bottom: Dipolar XY systems without pretraining. We show the energy error per site for $6\times 6$ AFM (left) and FM (right) dipolar XY systems after $2000$ and $5000$ epochs of only energy driven training. } 
\label{fig:VMCresults}    
\end{figure}


In the second part of the hybrid training, we use VMC, estimating the expectation value of the energy $\langle E\rangle $ of the TQS using Eq. \eqref{eq:E}. In order to stabilize the training, see e.g. Ref. \cite{hibat-allah_recurrent_2020}, we use
\begin{align}
    \mathcal{C} = \sum_{\vec{\sigma}}\vert \psi_{\vec{\theta}}(\sigma)\vert ^ 2 \left[  E^{\mathrm{loc}}_{\vec{\theta}} (\sigma)-\langle E^{\mathrm{loc}}_{\vec{\theta}}\rangle\right]
    \label{eq:Cost}
\end{align}
to minimize both the local energy as well as the variance of the gradients.

As explained in the main text, the optimization of Eq. \eqref{eq:Cost} is in general very complicated due to the very rugged loss landscape with many local minima \cite{Bukov2021}. Furthermore, the training of the autoregressive transformer is further complicated by the fact that stochastic reconfiguration \cite{Sorella1998,Sorella2001,becca_sorella_2017} and variants as used in Refs. \cite{rende_simple_2023,chen2023efficient} become unstable for autoregressive architectures \cite{donatella2023dynamics,lange2023neural}. Hence, we use the Adam optimizer \cite{kingma2017adam} to optimize according to Eq. \eqref{eq:Cost}.  \\

The results for an only energy based training for $6\times 6$ TFIM and dipolar XY systems are shown in Fig.~\ref{fig:VMCresults} (top and bottom, respectively) for antiferromagnetic (AFM) and ferromagnetic exchange interactions. It can be seen that the FM case is in general easier to learn for the TQS, reaching energy errors below $1\%$ after only $1000$ ($2000$) training steps for the TFIM (dipolar XY model). Hence, we do not consider these cases for the hybrid training scheme. For the AFM case, the training becomes more challenging, yielding longer convergence times for most parameter regimes.

\FloatBarrier
\section{Additional results \label{appendix:AddRes}}
In this section, we provide additional results obtained using the hybrid training scheme for the 2D TFIM \eqref{eq:HamTFIM} and the 2D dipolar XY model \eqref{eq:Ham}. Furthermore, we present results on an experimentally realized TFIM system \cite{Scholl2021}, for which experimental data from a single basis is available.

\subsection{Additional results on the 2D dipolar XY model \label{appendix:AddDipXY}}

\subsubsection{$6\times 6$ systems}
\begin{figure*}[t]
\centering
\includegraphics[width=0.95\textwidth]{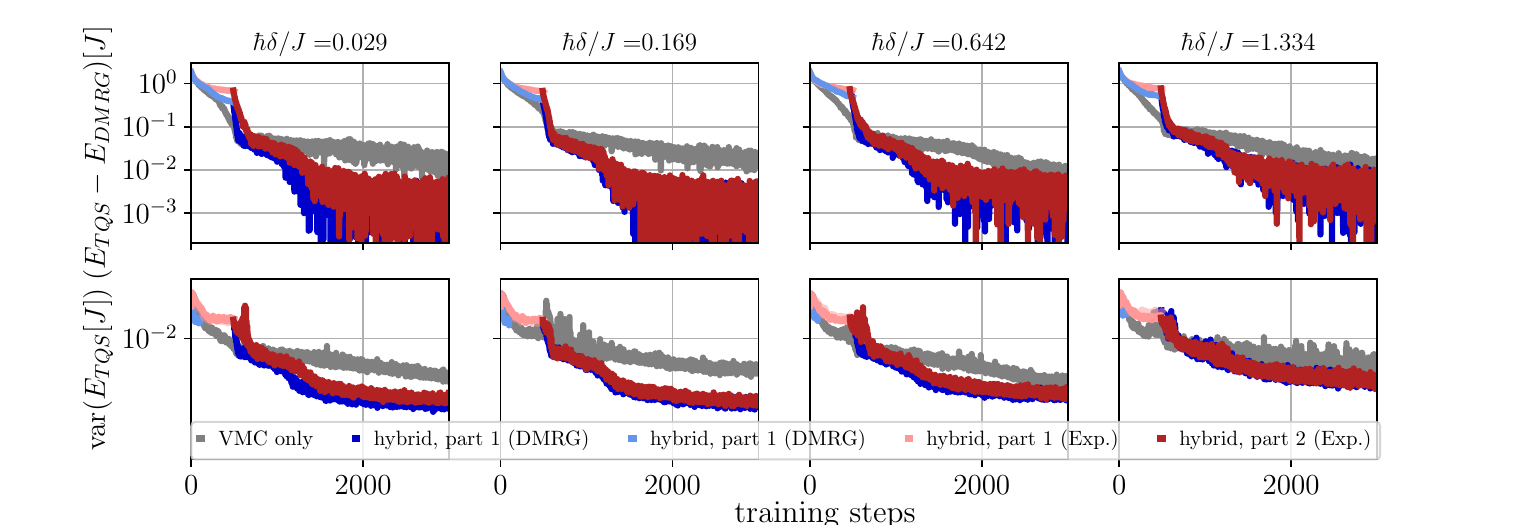}
\caption{ Training curves for $6\times 6$ dipolar XY systems for different $\delta$, with only energy based training (VMC, gray lines) and hybrid training (red lines) with numerical (light red lines) and experimental (solid red lines) data. Top: Energy error per site. Bottom: Variance of energies during the training. } 
\label{fig:TrainingCurves6x6}    
\end{figure*}

\begin{figure}[t]
\centering
\includegraphics[width=0.95\textwidth]{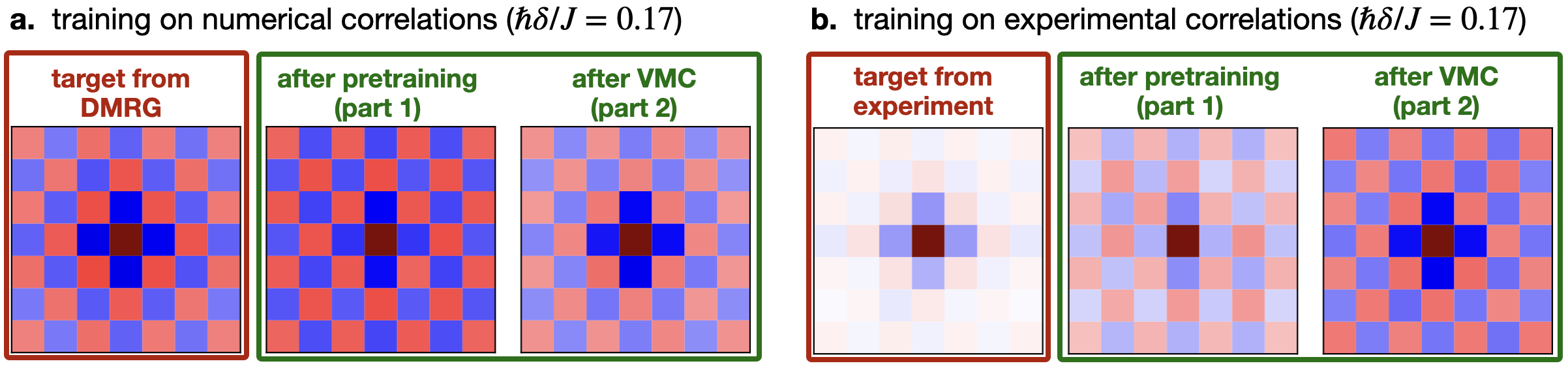}
\caption{Correlation maps for the dipolar XY model on the $6\times 6$ lattice. We show $\langle \sigma^x_\mathbf{i}\sigma^x_\mathbf{j}\rangle $, with $\mathbf{i}$ located in the middle of the system, for $\hbar \delta/J=0.17$. During the pretraining, we train on target correlations $\langle \sigma^x_i\sigma^x_j\rangle $ from numerics (experiment), see \textbf{a.} (\textbf{b.}) on the left, leading to a parameterization of the TQS that approximates these correlations (middle). The correlation maps of the TQS after the complete hybrid training including the data-driven second part are shown on the right. Note that the correlations after the pretraining can have higher values than the target, since we employ a scaled MSE loss that penalizes larger values than the target less than smaller values. The correlation maps for the TQS can show a small asymmetry between vertical and horizontal directions since we only apply $180$ degree rotational symmetry. The asymmetry vanishes upon convergence (see \textbf{a.} and \textbf{b.} on the right).} 
\label{fig:6x6_additional}    
\end{figure}

\begin{figure}[t]
\centering
\includegraphics[width=0.8\textwidth]{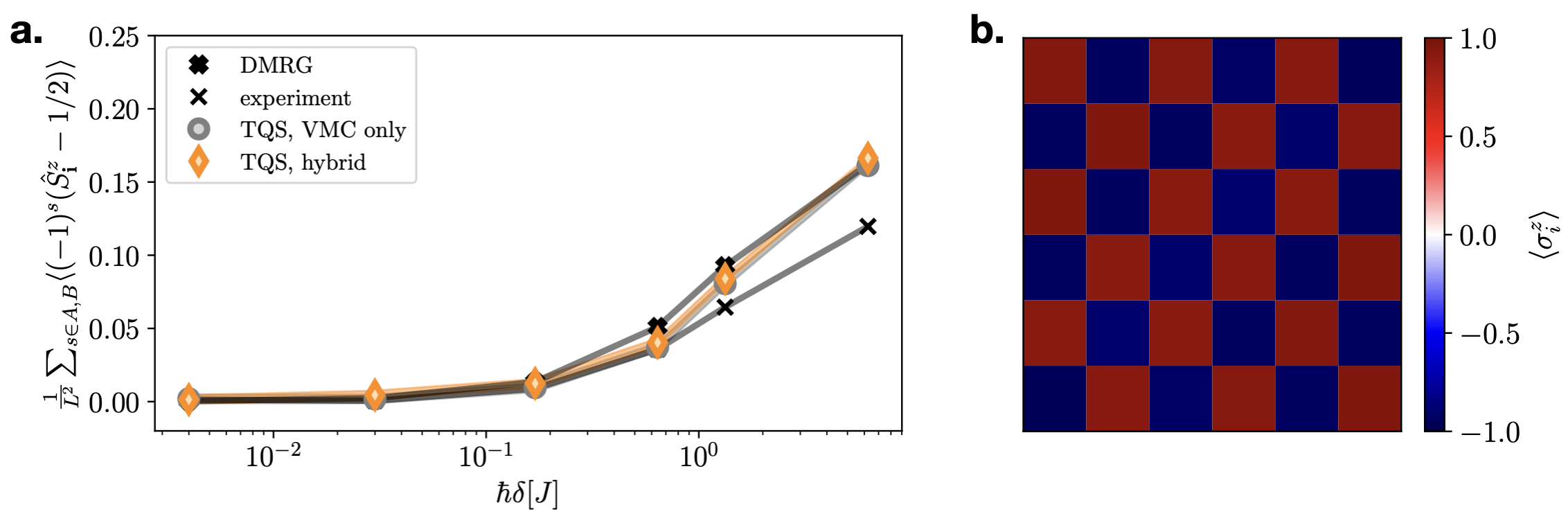}
\caption{Magnetization in $Z$ direction for the dipolar XY model on the $6\times 6$ lattice. \textbf{a.} We show $\frac{1}{L^2}\sum_{s\in A,B}\langle (-1)^s (\hat{S}^z_\mathbf{i}-1/2)\rangle $ for different light shifts $\delta$, for DMRG (black thick crosses), experiment (thin black crosses), TQS without pretraining (gray circles) and TQS with pretraining on experimental data (orange diamonds). \textbf{b.} The local magnetization $\langle\sigma^z_\mathbf{i}\rangle$ for $\hbar \delta=6.24$ obtained from the pretrained TQS. In both figures, we use the KL divergence for snapshots from the computational basis and the spin correlation maps from the $X$ basis for the pretraining. We use $10,000$ samples for each data point.} 
\label{fig:6x6_staggeredmag}    
\end{figure}
In Fig.~\ref{fig:TrainingCurves6x6} we show exemplary training curves for the $6\times 6$ systems and different Hamiltonian parameters $\delta$ up to epoch $3000$. It can be seen that both energy and the variance of the energies are decreased when using the hybrid scheme, both for experimental (light and dark red lines) and numerical (light and dark blue lines) data. \\

Furthermore, we show the spin-spin correlation maps for exemplary $\hbar \delta/J=0.17$ from the numerical (experimental) training data as well as from the TQS after pretraining and after the full hybrid scheme (left to right) in Fig.~\ref{fig:6x6_additional} \textbf{a} (\textbf{b}). After the first stage, the TQS captures already the AFM correlations. Note that for both DMRG and experimental data, the spin-spin correlation are slightly stronger than for the target, which is due to the asymmetric MSE \ref{eq:scaledMSE} that forces the TQS to represent stronger correlations rather than smaller correlations than the target. The final result obtained from the TQS agrees very well with the DMRG result in both cases. Furthermore, after the pretraining the correlation maps for the TQS can show a small asymmetry between vertical and horizontal directions since we only apply $180$ degree rotational symmetry (see middle plots in Fig.~\ref{fig:6x6_additional} \textbf{a} and \textbf{b}). The asymmetry vanishes upon convergence (see Fig.~\ref{fig:6x6_additional} \textbf{a} and \textbf{b} on the right).\\

Finally, Fig.~\ref{fig:6x6_staggeredmag} shows the magnetization in the computational basis. Fig.~\ref{fig:6x6_staggeredmag}\textbf{a} reveals good agreement of the staggered magnetization
\begin{align}
    m_\mathrm{stag} = \frac{1}{L^2}\sum_{s\in A,B}\langle (-1)^s (\hat{S}^z_\mathbf{i}-1/2)\rangle 
    \label{eq:staggeredmag}
\end{align}
for the TQS (with and without pretraining) with the DMRG results. The staggered magnetization obtained from the experiment deviates only slightly for large light shifts $\delta$. Fig.~\ref{fig:6x6_staggeredmag}\textbf{b} shows the local magnetization $\langle\sigma^z_\mathbf{i}\rangle$ for $\hbar \delta=6.24$.

\subsubsection{$10\times 10$ systems}
\begin{figure*}[t]
\centering
\includegraphics[width=0.9\textwidth]{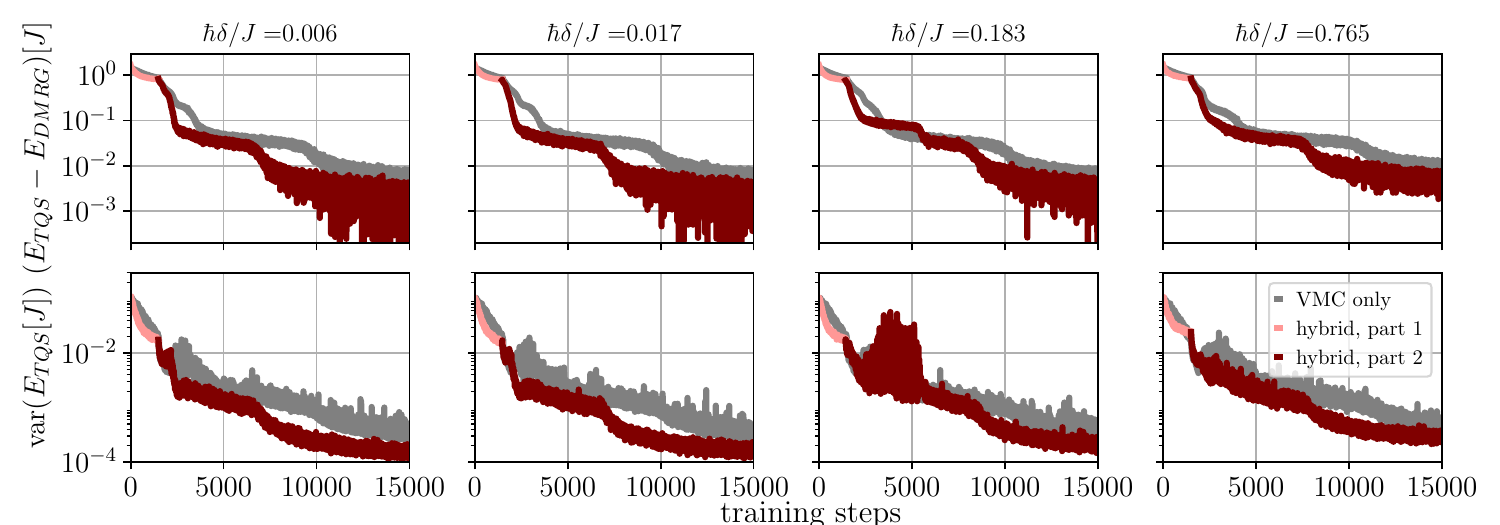}
\caption{Training curves for $10\times 10$ dipolar XY systems for different $\delta$, with only energy based training (VMC, gray lines) and hybrid training (red lines) using experimental data. Top: Energy error per site. Bottom: Variance of energies during the training. } 
\label{fig:TrainingCurves_10x10}    
\end{figure*}

In Fig.~\ref{fig:TrainingCurves_10x10} we show exemplary training curves for the $10\times 10$ systems and different Hamiltonian parameters $\delta$ up to epoch $15000$. Again, both energy and the variance of the energies are decreased when using the hybrid scheme, here for experimental data(light and dark red lines), compared to only using VMC (gray lines).\\

\begin{figure*}[htp]
\centering
\includegraphics[width=0.8\textwidth]{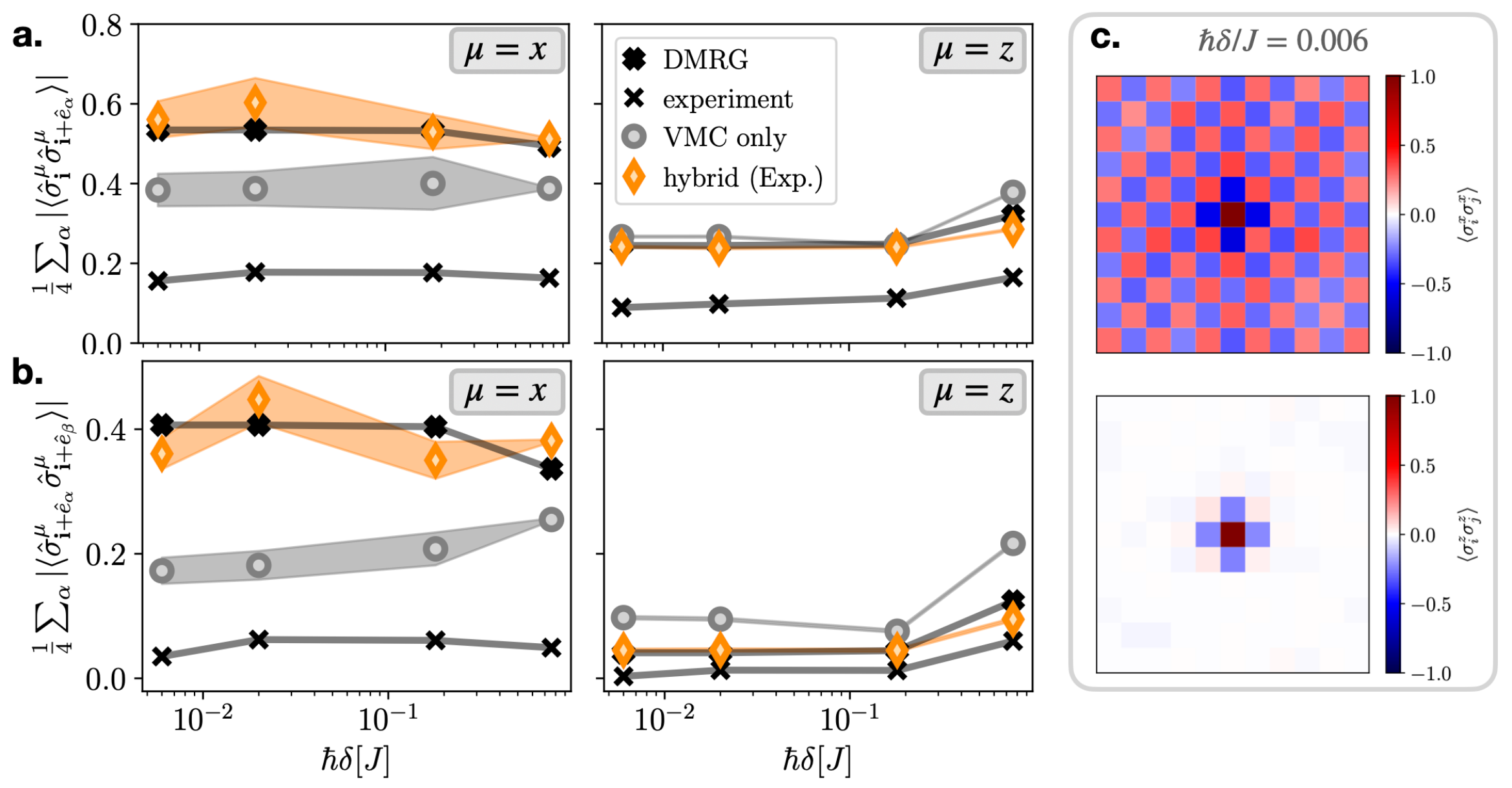}
\caption{Spin-spin correlations for $10\times 10$ dipolar XY model systems: The nearest-neighbor spin correlations (\textbf{a.}) and next-nearest-neighbor spin correlations (\textbf{b.}) averaged over all neighbors for $\mu=x$ (left) and $\mu=z$ (right), from DMRG and experiment (black) as well as the TQS without and with pretraining (with the Wasserstein and MSE loss) using experimental data (gray and orange). We use $40,000$ samples for each data point. Errorbars denote the standard deviation of the mean. \textbf{c.} Spin-spin correlation maps for fixed site $\mathbf{i}=(5,5)$ in the middle of the system. } 
\label{fig:Corrs10x10}    
\end{figure*}

Furthermore, the spin-spin correlations for $10\times 10$ dipolar XY model systems are shown in Fig.~\ref{fig:Corrs10x10}. We plot both nearest-neighbor spin correlations (Fig.~\ref{fig:Corrs10x10}\textbf{a}) and next-nearest-neigbor spin correlations (Fig.~\ref{fig:Corrs10x10}\textbf{b}) in the $X$ (left) and $Z$ (right) basis. In all cases, the results from DMRG and experiment are denoted as black thick and thin crosses respectively. The TQS results without and with pretraining using experimental data are shown in gray and orange. Similar to the $6\times 6$ systems in Fig.~\ref{fig:6x6}, the TQS results agree well with the DMRG results when using the hybrid scheme. Note that this is remarkable since we pretrain on the experimental data, which, as can be seen in Fig.~\ref{fig:Corrs10x10} for the back thing crosses, under-estimates both nearest-neighbor correlations. Without pretraining, the TQS shows smaller nearest-neighbor and next-nearest-neighbor correlations than DMRG for the $X$ basis, while slightly overestimating both correlations in the $Z$ basis. Spin-spin correlations for a fixed site $\mathbf{i}=(5,5)$ in the middle of the system in $X$ (top) and $Z$ (bottom) basis are shown in Fig.~\ref{fig:Corrs10x10}\textbf{c}.

\begin{figure}[htp]
\centering
\includegraphics[width=0.8\textwidth]{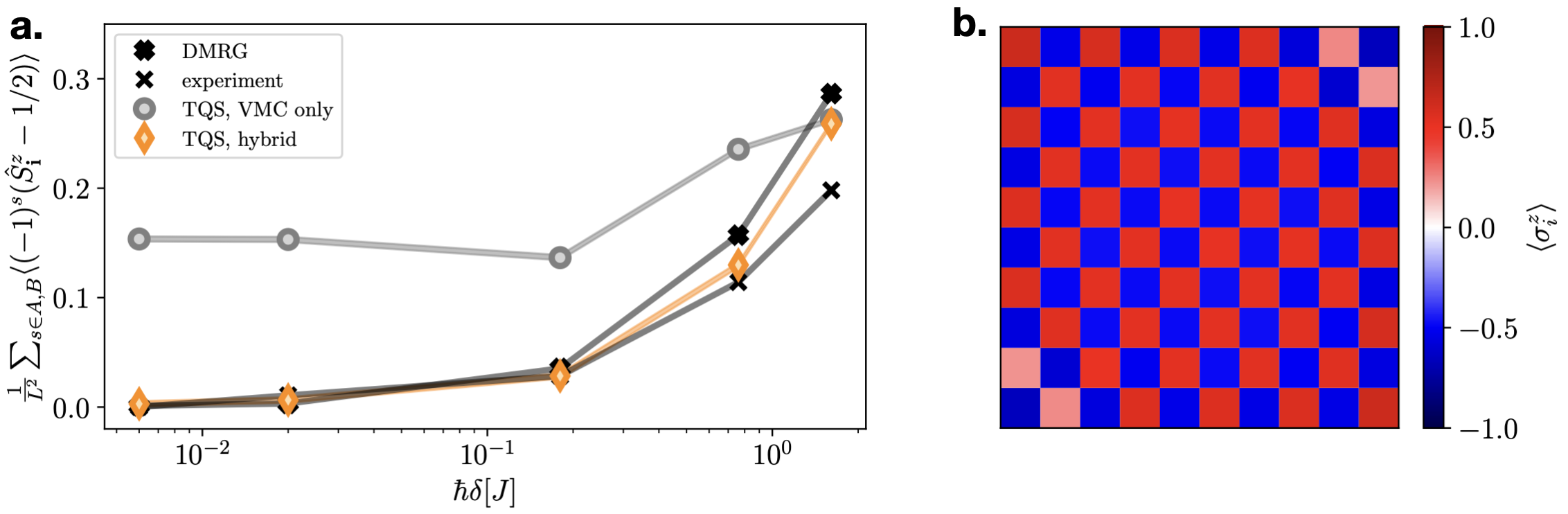}
\caption{Magnetization in $Z$ direction for the dipolar XY model on the $10\times 10$ lattice. \textbf{a.} We show $\frac{1}{L^2}\sum_{s\in A,B}\langle (-1)^s (\hat{S}^z_\mathbf{i}-1/2)\rangle $ for different light shifts $\delta$, for DMRG (black thick crosses), experiment (thin black crosses), TQS without pretraining (gray circles) and TQS with pretraining on experimental data (orange diamonds). \textbf{b.} The local magnetization $\langle \sigma^z_\mathbf{i}\rangle$ for $\hbar \delta=1.61$ obtained from the pretrained TQS. In both figures, we use the Wasserstein distance for snapshots from the computational basis and the spin correlation maps from the $X$ basis for the pretraining. We use $40,000$ samples for each data point.} 
\label{fig:10x10_staggeredmag}    
\end{figure}

Lastly, Fig.~\ref{fig:10x10_staggeredmag} shows the magnetization in the computational basis. Fig.~\ref{fig:10x10_staggeredmag}\textbf{a} reveals good agreement of the staggered magnetization \eqref{eq:staggeredmag}
for the pretrained TQS with the DMRG results. As for the smaller systems in Fig.~\ref{fig:6x6_staggeredmag}, the staggered magnetization obtained from the experiment deviates only slightly for the largest light shift $\hbar \delta=1.61$. However, in contrast to the $6\times 6$ systems, for the $10\times10$ systems the staggered magnetization is highly overestimated for $\hbar \delta <1$ when omitting the pretraining. Furthermore, Fig.~\ref{fig:6x6_staggeredmag}\textbf{b} shows a small inhomogeneity of the local magnetization $\langle\sigma^z_\mathbf{i}\rangle$ for $\hbar \delta=1.61$ at the lower left and upper right corners, resulting from the sampling path, where $2\times 2$ patches are sampled in a 1D manner, combined with a $180$ degree rotational symmetry. We expect that this feature vanishes upon applying more symmetries, e.g. translational symmetry.

\FloatBarrier
\subsection{Additional results on the 2D TFIM \label{appendix:AddTFIM}}

In Fig.~\ref{fig:6x6_additional} we show additional results for the 2D AFM TFIM on a $6\times 6$ lattice. Fig.~\ref{fig:6x6_additional}\textbf{a} shows the exemplary learning curve for $h/J=5.0$. It can be seen that while pretraining reaches an energy error per site of around $>0.1 J$, the second VMC part reduces the error below $0.001J$ below within less than $500$ optimization steps (Fig.~\ref{fig:6x6_additional}\textbf{a} left). Without pretraining, the optimization is stuck at around $0.1J$ for almost $1000$ epochs before decreasing to a slightly higher energy than with pretraining. Furthermore, the variance is highly decreased when using pretraining (see Fig.~\ref{fig:6x6_additional}\textbf{a} right).

In Fig.~\ref{fig:6x6_additional}\textbf{b}, we additionally show the average magnetization $\langle \hat{S}^\mu_\mathrm{tot}\rangle$ in $\mu=x,z$ directions after pretraining (part $1$) and VMC (part $2$) staged of the hybrid training. In both cases, the average magnetization agrees well with the DMRG results. In particular, the increase of magnetization in the $X$ direction is correctly captured already in the pretraining.

\begin{figure}[htp]
\centering
\includegraphics[width=0.95\textwidth]{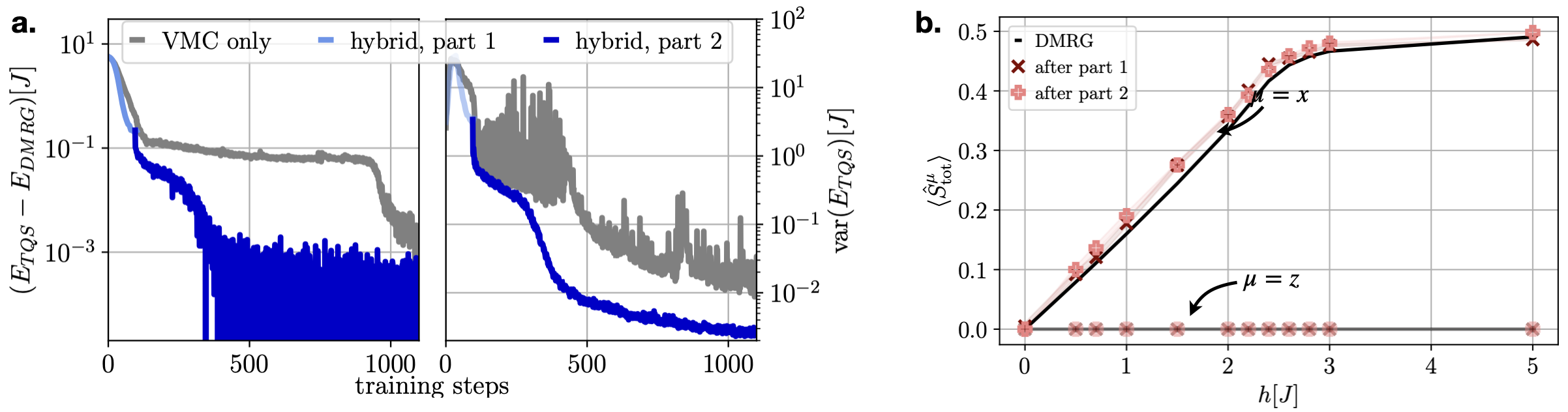}
\caption{Additional results on the 2D AFM TFIM on a $6\times 6$ lattice: \textbf{a.} Exemplary learning curve for $h/J=5.0$. \textbf{b.} The average magnetization $\langle \hat{S}^\mu_\mathrm{tot}\rangle$ in $\mu=x,z$ directions after part $1$ and part $2$ of the hybrid training. We use $10,000$ samples for the evaluation of $\langle \hat{S}^\mu_\mathrm{tot}\rangle$ and errorbars denote the standard deviation of the mean.} 
\label{fig:6x6TFIM_additional}    
\end{figure}
\FloatBarrier

\subsection{Experimental realization of the 2D TFIM \label{appendix:TFIMExp}}
\subsection{Training data}
For the computational $Z$ basis, we use the experimentally available data from Ref. \cite{Scholl2021}, see Appendix~\ref{appendix:Data}. For comparison, we train the transformer wave function with numerical data from the same basis. Furthermore, we use the numerically obtained locally resolved magnetization from $X$ basis for the pretraining.

\subsection{Results}

\begin{figure*}[t]
\centering
\includegraphics[width=0.95\textwidth]{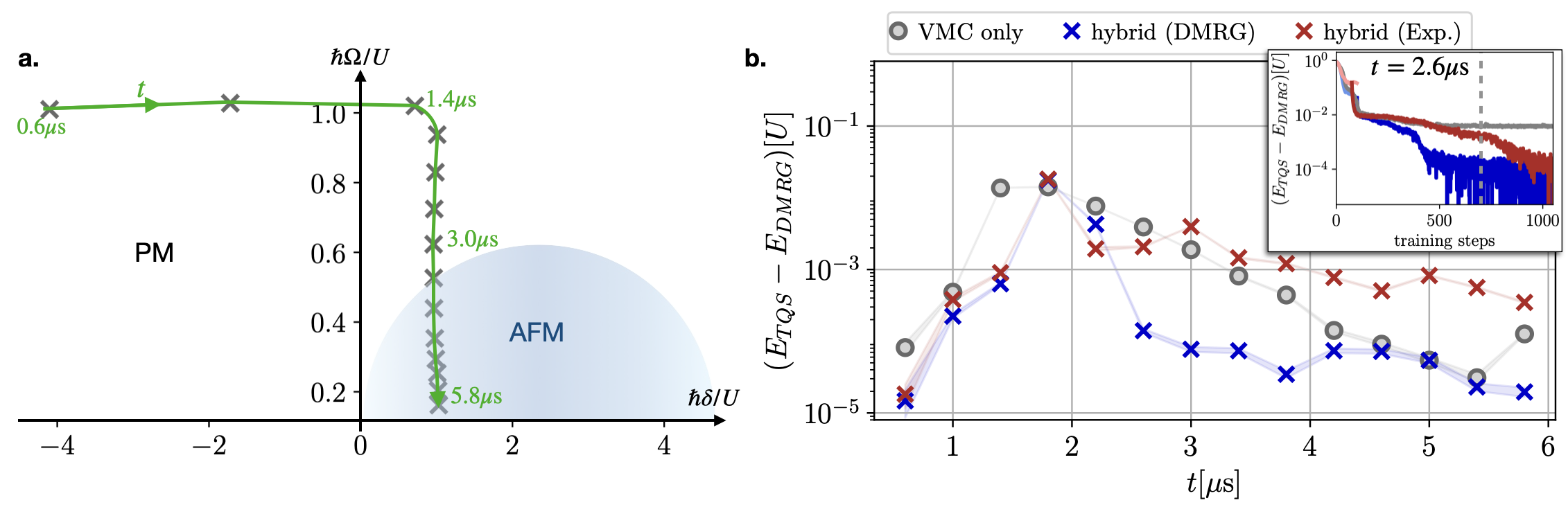}
\caption{Results for the experimentally realized TFIM \eqref{eq:H_TFIM} on a $10\times 10$ lattice. \textbf{a.} Experimentally realized values for $\Omega(t)$ and $\delta(t)$ over time $t$. During this sweep, the system is transferred from a initial paramagnetic (PM) ground state $\vert \downarrow\downarrow\dots \downarrow\rangle$ to an antiferromagnetic (AFM) phase denoted in blue. \textbf{b.} Energy error for the Hamiltonians realized at time $t$ without pretraining (gray), with pretraining on numerical data from both $Z$ and $X$ bases (blue) and with pretraining on experimental data from the $Z$ and numerical data from the $X$ basis (red) after $700$ optimization steps. } 
\label{fig:10x10TFIMExp}    
\end{figure*}

In this section, we present results on the experimentally realized TFIM, see \eqref{eq:H_TFIM}, on a $10\times 10$ lattice. Fig.~\ref{fig:10x10TFIMExp}\textbf{a} shows the values for $\Omega(t)$ and $\delta(t)$ realized in the experiment \cite{Scholl2021} over the sweep time $t$. During this sweep, the system is transferred from a initial paramagnetic (PM) ground state $\vert \downarrow\downarrow\dots \downarrow\rangle$ to an antiferromagnetic (AFM) phase $\vert \downarrow\uparrow\downarrow\dots \downarrow\rangle$ \cite{Scholl2021}.

The respective ground state energy errors per site for $\Omega(t)$ and $\delta(t)$ during the sweep are shown in Fig.~\ref{fig:10x10TFIMExp}\textbf{b}, and the hyperparameters that were used during the training are listed in Tab.~\ref{tab:paramsTFIMExp}. In Fig.~\ref{fig:10x10TFIMExp}\textbf{b}, gray lines denote the result without pretraining, blue and red lines the results with pretraining on numerical data from both $Z$ and $X$ bases and with pretraining on experimental data from the $Z$ and numerical data from the $X$ basis respectively, all of them evaluated after only $700$ training iterations. All errors with and without pretraining are already very low in this early stage of the optimization: Except for intermediate times $1.5\mu\mathrm{s}<t<3\mu\mathrm{s}$, all energy errors are on the order or below $0.1\%$.

When pretraining with numerical data, the energy error per site decreases w.r.t. the only VMC based training, in particular for $1.8\mu\mathrm{s}<t<4\mu\mathrm{s}$, i.e. in the regime where a phase transition between PM and AFM phase is expected, the error decreases by more than an order of magnitude. For other times $t$, the energy error per site is slightly decreased or the same ($t=1.8\mu\mathrm{s}$). For the experimental data however, only a slight improvement ($t<3\mu\mathrm{s}$) or even a higher error are observed ($t\geq3\mu\mathrm{s}$). Note that in the latter regime the results obtained with only VMC based training are already below $<0.1\%$, indicating that in the case of such fast convergence, the hybrid scheme when using imperfect training data can yield slower convergence.

\begin{table}[t]
\centering
\begin{tabular}{|l|c|}\hline
 & $10\times 10$  \\\hline
number of transformer layers & $2$  \\
hidden dimension $d_h$ & $32$  \\
FFNN dimension (see Fig.~\ref{fig:Transformer}) & $4d_h$  \\
total number of parameters & $28,800$  \\
pretraining learning rate & $1\cdot 10^{-3}$   \\
VMC learning rate $l_\mathrm{max}$ & $ 2\cdot  10^{-3}$\footnote{Except for $t=1.8\mu \mathrm{s}$, where the pretraining learning rate is decreased by a factor $1/5$ to avoid getting stuck in a local minimum.}  \\
learning rate schedule $l(n)$ &  $n_\mathrm{start}=300$, $\gamma=0.9998$ \\
maximal no. of pretraining steps $n_\mathrm{pretrain}^\mathrm{max}$ & $100$\\
number of samples in VMC training (part 2) & $512$\\
number of samples used for the data points in Fig.~\ref{fig:10x10TFIMExp} & $100\times 512$\\
errorbars in Fig.~\ref{fig:10x10TFIMExp} & standard deviation of the mean\\
number of training steps used in Fig.~\ref{fig:10x10TFIMExp} & $700$\\\hline
\end{tabular}
\caption{Transformer parameters and training hyperparameters for the $10\times 10$ experimental TFIM systems \eqref{eq:H_TFIM} (Fig.~\ref{fig:10x10TFIMExp}). We refer to the total number of training steps including the pretraining as $n$. Futhermore, we use a exponential decay rate starting at epoch $n_\mathrm{start}$ with decay rate $\gamma$.}
\label{tab:paramsTFIMExp}
\end{table}

\section{Training data \label{appendix:Data}}
\subsection{Experimental data}

The experimental setup consists of two-dimensional arrays of $^{87}\text{Rb}$ atoms trapped in optical tweezers. The atoms are first loaded into the tweezers and assembled in square arrays of $N$ atoms with a lattice spacing $a$~\cite{Barredo2016}. We then cool down the atoms to $T\sim 20\,\mu$K, optically pump them in the hyperfine state $\ket{g} = \ket{5S_{1/2}, F = 2, m_F = 2}$ and switch off the tweezer lights. Starting from this point, two different configurations of the experimental setup allow us to implement either the dipolar XY model \cite{Chen2023} or the transverse field Ising one \cite{Scholl2021}.

\subsubsection{Dipolar XY model} 
To implement the XY model, we encode our spin states between two Rydberg states of opposite parities $\ket{\uparrow} = \ket{60S_{1/2},m_J = 1/2}$ and $\ket{\downarrow} = \ket{60P_{1/2},m_J = -1/2}$. Resonant dipole-dipole interactions between the spins naturally realize the dipolar XY model \eqref{eq:Ham}, which we state here again for convenience:
\begin{equation}
	\label{eq:H_XY}
	\hat{\cal H}_{\rm XY} =  {J\over 2}
	\sum_{i < j}  \frac{a^3}{r_{ij}^3}  (\hat{\sigma}^x_i \hat{\sigma}^x_j + \hat{\sigma}^y_i \hat{\sigma}^y_j) ~,
\end{equation} 
with $a = 12\,\mu$m and $J/h = -0.77\,$MHz~\cite{Leseleuc2017}. The spins can be manipulated using microwaves at $\sim 16.7\,$GHz coupling $\ket{\uparrow}$ to $\ket{\downarrow}$.
To prepare the ground state, we apply the following procedure. After having prepared all the spins in $\ket{g}$, we drive a two-photon
transition (STIRAP) to excite all atoms in $\ket{\uparrow}$. This step is realized using two laser beams of wavelengths $420\,$nm and $1013\,$nm that respectively couple $\ket{g}$ and $\ket{\uparrow}$ to the intermediate state $\ket{i} = \ket{6P_{3/2}, F = 3, m_F = 3}$. Then, we address half of the atoms in a staggered configuration using off-resonant $1013\,$nm beams to apply a local light-shift $\delta$ on these atoms. The total Hamiltonian thus reads $\hat{\cal H} = \hat{\cal H}_{\rm XY} + \hbar\delta\sum_i \hat{n}_i$ with $\hat{n}_i = 0$ for the non-addressed atoms and $\hat{n}_i = (1+\sigma^z)/2$ for the addressed atoms. Using these local light shifts in combination with microwave pulses, we prepare the system in a classical N\'eel state $\ket{\uparrow\downarrow\uparrow\downarrow\cdots}$ which is for $|\hbar\delta|\gg |J|$ the ground state of $\hat{\cal H}$ (see more details in Ref.~\cite{Chen2023, Chen2023a}). We then adiabatically decrease the applied light-shifts as $\delta(t) = \delta_0 e^{-t/\tau}$ (with $\delta_0/(2\pi) \approx 15\,$MHz and $\tau = 0.3\,\mu$s) thus connecting the N\'eel state to the XY antiferro-magnetic ground state. At the end of the ramp, we readout the state of the atoms. Using an on-resonant $1013\,$nm light beam, we transfer the $\ket{\uparrow}$ atoms to $\ket{i}$ from which they spontaneously decay in the ground state $(5S_{1/2})$ and turn back on the tweezer lights. The atoms transferred in $(5S_{1/2})$ are trapped and imaged, while the ones left in the Rydberg state are expelled from their traps. Thus, we map the $\ket{\uparrow}$ and $\ket{\downarrow}$ state to the presence or absence of the corresponding atom. Additionally, when we want to measure the spins along $x$, we rotate them by applying a
microwave $\pi/2$-pulse prior to the readout sequence.


In the experiment, also van der Waals interactions are present:
\begin{align}
    \hat{\mathcal{H}}_\mathrm{vdW} = \sum_{i<j} \frac{a^6}{r_{ij}^6}\left[U^{PP}_6 \hat{P}_i^\uparrow \hat{P}_j^\uparrow + U^{SS}_6 \hat{P}_i^\downarrow \hat{P}_j^\downarrow + U^{SP}_6 \hat{P}_i^\downarrow \hat{P}_j^\uparrow \right]
\end{align}
with the projection operators $\hat{P}_i^{\uparrow/\downarrow} =\frac{1}{2}\pm\hat{S}_i^z$, and $U^{PP}_6/h=-0.008\mathrm{MHz}$, $U^{SS}_6/h=0.037\mathrm{MHz}$ and $U^{SP}_6/h=-0.0007\mathrm{MHz}$.\\
Note that in the experiment, $10\times 10 $ and $6\times 7 $ sites systems were realized. Since the latter is not compatible with our $2\times 2$ patches, see Appendix~\ref{appendix:transformer}, we simulate a $6\times 6$ system and cut off samples and correlation maps for the pretraining.\\

\subsubsection{(Experimentally realized) Transverse field Ising model} To explore the Ising model, we encode our spin states in $\ket{\downarrow} = \ket{g}$ and the Rydberg state $\ket{\uparrow} = \ket{75S_{1/2}, m_J = 1/2}$ with van der Waals interactions of $U^{\uparrow\uparrow}/h \approx 1.95\,$MHz for $a = 10\,\mu$m. Using the two lasers at $420\,$nm and $1013\,$nm, we drive a two-photon transition, thus coupling the spin states with an effective Rabi frequency $\Omega$ and detuning $\delta$. The Hamiltonian reads:
\begin{equation}
	\label{eq:H_TFIM}
	\hat{\cal H}_{\rm Exp.TFIM}=U^{\uparrow\uparrow}
	\sum_{i < j}  \frac{a^6}{r_{ij}^6} \hat{n}^\uparrow_i \hat{n}^\uparrow_j + \frac{\hbar\Omega}{2} \sum_{i} \hat{\sigma}^x_i - \hbar\delta\sum_{i}\hat{n}^\uparrow_i~,
\end{equation}
with $\hat{{n}}^\uparrow=\frac{1}{2}(1+\hat{\sigma}^z_i)$ and $r_{ij}$ given in units of the lattice constant.
As explained in Sec.~\ref{appendix:TFIMExp} and in Ref.~\cite{Scholl2021}, to probe the antiferromagnetic ground state of the TFIM model, we sweep $\Omega$ and $\delta$ to transfer the system from its initial state $\ket{\downarrow\downarrow\downarrow\downarrow\cdots}$ to the Ising antiferromagnetic ground state. At the end of the sequence, we readout the state of the system by switching back on the tweezer light. The atoms in $\ket{\downarrow} = \ket{g}$ are recaptured and imaged, while the ones that stayed in the Rydberg state $\ket{\uparrow}$ are lost.

\subsection{Numerical data from DMRG}
For the training with numerical data we use the single-site density matrix renormalization group (DMRG) algorithm \cite{SCHOLLWOCK201196} implemented in the package SyTen \cite{syten1,syten2}. \\

\begin{itemize}
    \item \textit{Dipolar XY model.} For the 2D dipolar XY model \eqref{eq:Ham}, we explicitly employ $U(1)$ symmetry and use bond dimensions up to $\chi_\mathrm{max}=2048$ both for the $10\times 10$ and $6\times 6$) systems, a tolerance of $10^{-5}$ and a truncation threshold of $10^{-11}$. Further numerical analysis of the dipolar XY model can be found in Ref. \cite{Chen2023}. 
    \item \textit{Transverse field Ising model.} For the 2D TFIM \eqref{eq:HamTFIM} on the $6\times 6$ lattice, we use the same parameters with $\chi_\mathrm{max}=512$.
    \item \textit{(Experimentally realized) Transverse field Ising model.} For the experimentally realized TFIM \eqref{eq:H_TFIM}, we use bond dimensions up to $\chi_\mathrm{max}=1024$ for the $10\times 10$ systems, a tolerance of $10^{-5}$ and a truncation threshold of $10^{-11}$. The long-range interactions in Eq. \eqref{eq:H_TFIM} are cut off for $r>4$. 
\end{itemize}

\end{document}